\documentclass[showpacs,showkeys,preprintnumbers,amsmath,amssymb,twocolumn,pre,floatfix]{revtex4}

\usepackage{amssymb,amsfonts}
\usepackage{amsmath}
\usepackage{graphicx}
\graphicspath{{Figures/}}
\begin{document}

\title{The Role of Protein Fluctuation Correlations in Electron Transfer in Photosynthetic Complexes}

\author{Alexander I. Nesterov}%
   \email{nesterov@cencar.udg.mx}
\affiliation{Departamento de F{\'\i}sica, CUCEI, Universidad de Guadalajara,
Av. Revoluci\'on 1500, Guadalajara, CP 44420, Jalisco, M\'exico}

\author{Gennady P.  Berman}
 \email{gpb@lanl.gov}
\affiliation{ Theoretical Division, T-4, Los Alamos National Laboratory, and the New
Mexico Consortium,  Los Alamos, NM 87544, USA}

\date{\today}

\begin{abstract}
We consider the dependence of the electron transfer in photosynthetic complexes on  correlation properties of random fluctuations of the protein environment. The electron subsystem is modeled by a finite network of connected electron (exciton) sites. The fluctuations of the protein environment are modeled by random telegraph processes, which act  either collectively (correlated) or independently (uncorrelated) on the electron sites. We derived an exact closed system of first-order linear differential equations with constant coefficients, for the average density matrix elements and for their first moments. Under some conditions, we obtain analytic expressions for the electron transfer rates. We compare the correlated and uncorrelated regimes, and demonstrated numerically that the uncorrelated fluctuations of the protein environment can, under some conditions, either  increase or decrease the electron transfer rates.

\end{abstract}



\preprint{LA-UR-14-29075}

\maketitle


\section{ Introduction}
   In a photosynthetic organism, sunlight is absorbed in the light-harvesting complex (LHC) or antenna,  by a light-sensitive (chlorophyll or carotenoid) molecule. This is the first step in transforming solar energy into electron energy in the form of the exciton. This exciton travels through many connected sites (pigments) of the antenna complex, and finally reaches the reaction center (RC), where charge separation and chemical reactions take place.  (See, for example, \cite{1,2}, and  references therein.) The timescale of the primary processes of electron (exciton) transfer (ET) and charge separation are very fast, $t_{\rm prime}\approx 1-3\rm ps$.  There are two major theoretical challenges in describing these primary processes. The first problem is that the constant of interaction, $\lambda_n$,   between the electron site, $n$,  and the protein environment is usually not small. Indeed, the well-known Marcus formula for  the ET rate, $k_{da}$,  between the donor and the acceptor,  under the influence of the collective  protein thermal fluctuations, has the form \cite{3,4}: $k_{da}=(2\pi|V_{da}|^2/\sqrt{4\pi\varepsilon_rT})\exp[-(\varepsilon-\varepsilon_r)^2/4k_BT\varepsilon_r]$. Here $\varepsilon$  is the difference between the donor and the acceptor site energies; $V_{da}$  is the matrix element of the donor-acceptor interaction, $T$  is the absolute temperature;  $k_B$  is the Boltzmann constant; and $\varepsilon_r$  is the so-called reconstruction energy, $\varepsilon_r\propto\lambda^2$.  As one can see, the interaction constant, $\lambda$,  occurs in the  denominators of both the pre-exponential factor and in the exponent. This result cannot be obtained by using standard  perturbative approaches by expanding the initial expressions in a power series in the interaction constant, $\lambda$.  It is known that the Marcus formula is derived in the high-temperature limit. But the complete set of conditions for the applicability of the Marcus-type formula for complex biological networks with many sites is not known. Then, the first mathematical problem of dealing with ET in photosynthetic complexes is that there does not exist a closed system of equations (exact or approximate, but derived under controlled conditions) to deal with the ET processes. Indeed, the derivation of the equation for the reduced density matrix requires an averaging over the variables of the noisy protein environment. But (i) this procedure cannot be performed exactly and (ii)  controlled perturbation approaches do not exist. The second problem is related to a large number of the electron sites (or degrees of freedom) in the LHC and in the RC. This results in a multi-scale ET dynamics, so adequate coarse-grained procedures must be used  \cite{2,5}.

   In this situation, it would be useful to introduce a quantum exactly solvable (at least, numerically) model, which (i) applies to the ET in the photosynthetic complexes and (ii) does not include the above mentioned restrictions on the interaction constants. This model was introduced in our earlier publications \cite{6,7}, and used to describe (i) multi-scale ET dynamics and (ii) nonphotochemical quenching by a charge transfer state. 
   
   In this paper, we extend our approach \cite{6,7} to the case in which the random protein environment can act both collectively and independently  on all light-sensitive electron sites (pigments). Namely, for all  electron sites, we introduce both collective (correlated) and independent (uncorrelated) protein fluctuations, modeled by random telegraph processes. Protein environments modeled by random processes and by the thermal bath usually produce different long-time asymptotic behavior for the ET dynamics,  we consider that our approach is appropriate for the problems under consideration. Note also that protein environments in living organisms, have both noisy and thermal components \cite{1,8,9,10,11,12,13,14,15,16}. Uncorrelated protein fluctuations can act on their neighboring pigments as well as on other pigments in the real photosynthetic organisms. Experimentally, this can be verified by measuring the corresponding correlation functions of the protein fluctuations between different donor-acceptor sites. The correlation properties of protein fluctuations at different electron sites can also be modeled and simulated numerically using the standard molecular dynamics (MD) approaches. The main results of our paper include:
   \begin{itemize}
\item For rather  arbitrary photosynthetic complexes, we derived an exact closed system of first-order linear differential equations, with constant coefficients, for the averaged density matrix elements and for their first moments, which describe the quantum ET dynamics. 
\item We applied our model to determine the quantum ET dynamics of the simplest donor-acceptor system. Under some conditions, we derived analytic expressions for the ET rates, that are a generalization of the Marcus-type expression for noisy protein environment.  
\item We demonstrated numerically that the uncorrelated fluctuations of the protein environment can, under some conditions, either  increase or decrease the ET rates.
\item We compared our exact solutions with the corresponding approximate solutions, and found the conditions of the applicability of our perturbation approach. 
 \end{itemize}
The structure of the paper is the following. In Section II, we describe our model, and derive the closed system of differential equations for the averaged density matrix elements and for their moments. In Section III, we apply our approach to a specific ``donor-acceptor" system,  introduce the characteristic  parameters, and present the results of the numerical simulations for both exact and approximate solutions. In the Conclusion, we summarize our results and formulate some challenges for future research. In the Supplementary Material (SM), we present mathematical details of our approach, and additional illustrations on the action of correlated and uncorrelated protein environment on the ET.

\section{Description  of the model}

Consider a quantum system which is described by a time-dependent Hamiltonian, ${\mathcal H(t)}$.
We  assume that this Hamiltonian depends on some control parameters, $\lambda_a$. The noise associated with  fluctuations of these parameters is described by the functions,  $\delta\lambda_a(t)$, that depend on the random variables, $\xi_a(t)$. Expanding  the Hamiltonian to first order in $\xi_a(t)$, we have,
\begin{align} \label{Neq1}
{\mathcal H}(t) = {\mathcal H}_0 + \sum_a{\mathcal V}_a\xi_a(t),
\end{align}
where, ${\mathcal H}_0$, is the Hamiltonian of the system under consideration, and $\mathcal V_a$ is a matrix that describes the interaction with noise.
Using (\ref{Neq1}), we obtain the following equations of motion for the density matrix ($\hbar=1$),
\begin{align}\label{Meq2a}
\frac{d\rho}{dt} =i[\rho,{\mathcal H}_0 ] +
i [\rho, \sum_a \mathcal V_a,\xi_a(t)].
\end{align}
For the density matrix averaged over noise this yields,
\begin{align}\label{A3a}
\frac{d\langle\rho\rangle}{dt} =i [\langle\rho\rangle,{\mathcal H}_0] 
 +i\sum_a[\langle \rho  \xi_a(t)\rangle,{\mathcal V}_a ]  ,
  \end{align}
where the average, $\langle ...{\;}\rangle$, is taken over the random processes.

To close this system of differential equations (\ref{A3a}), we assume that the  fluctuations are produced by the independent random telegraph processes (RTPs),
\begin{align}\label{chi_8}
&\langle \xi_a(t)\rangle =0, \\
 &\langle \xi_a(t)\xi_b(t')\rangle = \delta_{ab}\sigma_a^2e^{-2\gamma_a \tau}.
 \end{align}

Employing the differential formula for the RTP \cite{KV2},
\begin{align}
&\Big(\frac{d}{dt} +2\gamma_a \Big)\langle  {\xi_a}(t)R[t;\xi_a(\tau) ]
\nonumber
\rangle = \\
&\Big\langle  {\xi}_a(t)\frac{d}{dt}R[t;\xi_a(\tau)]
\Big\rangle ,
\end{align}
where, $R[t;\xi_a(\tau)]$, is an arbitrary functional, we obtain from Eq. (\ref{A3a}) the following closed system of differential equations:
\begin{align}\label{A4a}
\frac{d\langle\rho\rangle}{dt} =&i [\langle\rho\rangle,{\mathcal H}_0] 
 +i\sum_a \sigma_a[\langle  X_a \rangle, {\mathcal V}_a ]   , \\
 \frac{d\langle X_a \rangle}{dt} =&i [\langle X_a\rangle,{\mathcal H}_0] +i \sigma_a[\langle  \rho \rangle, {\mathcal V}_a]
  \nonumber\\
&+i\sum_{b\neq a} \sigma_b[\langle  X_{ab} \rangle, {\mathcal V}_b]   - 2\gamma_a \langle X_a\rangle , \\
 \frac{d\langle X_{ab} \rangle}{dt} =&i [\langle X_{ab}\rangle,{\mathcal H}_0] 
 +i\sigma_a[\langle  X_{b} \rangle, {\mathcal V}_b] \nonumber \\
& + i\sigma_b[\langle  X_{a} \rangle, {\mathcal V}_b]   - 2(\gamma_a  + \gamma_b)\langle X_{ab}\rangle,
 \label{A4b}
\end{align}
where   $\langle  X_a(t)\rangle= \langle { \xi}_a(t) \rho(t)\rangle /\sigma_a$, and  
$\langle  X_{ab}(t)\rangle= \langle { \xi_a(t) \xi}_b(t) \rho(t)\rangle/(\sigma_a \sigma_b)$ ($a \neq b$). Note, that by using the properties of the RTP, one can show that, $\langle  X_{aa}(t)\rangle = \langle  \rho(t)\rangle$. Therefore, the diagonal elements of the matrix, $\langle  X_{ab}(t)\rangle$, do not add new equations to the system (\ref{A4a}) - (\ref{A4b}).

In the rest of this paper, we use Eqs. (\ref{A4a}) --  (\ref{A4b}) to study the two-level ``donor-acceptor" system (TLS) embedded in a noisy protein environment. We assume that two uncorrelated RTPs (generally, with different interaction constants)  act on both the donor and the acceptor. 

\begin{figure}[tbh]
	\scalebox{0.3}{\includegraphics{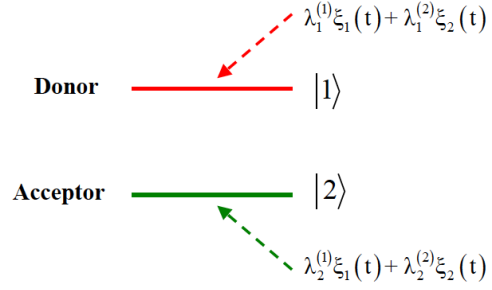}}
	\caption{(Color online) The two-level ``donor-acceptor" system (TLS)  interacting with two uncorrelated noisy environments, $\xi_1(t)$ and $\xi_2(t)$; $\lambda_n^{(a)}$ are the constants of interaction. The superscript, $a=1,2$, indicates the noisy environment, and the subscript, $n=1,2$, indicates the electron site.
		\label{TLS}}
\end{figure}

\section{Two-level ``donor-acceptor" system}

For a simplicity of consideration, we apply our approach to the TLS,  with the following Hamiltonian:
\begin{align} \label{Eq2b}
 \tilde{\mathcal H}= &\sum_n \varepsilon_n |n\rangle\langle  n | + \sum_{m \neq n} V_{mn} |m\rangle\langle  n | \nonumber\\
 &+ \sum_{m,n} \lambda_{mn}(t)|m\rangle\langle  n |,
 \quad m,n = 0,1,
\end{align}
where the functions, $\lambda_{mn}(t)$, describes the influence of noise. When the matrix elements, $V_{nm}$, are absent, the diagonal matrix elements, $\lambda_{nn}(t)$, are responsible for decoherence -- the decay of the non-diagonal density matrix elements. When $\lambda_{mn}(t)=0$ ($m\neq n$), relaxation in the system occurs only if $V_{mn}\ne 0$.  When $V_{mn}= 0$, the off-diagonal matrix elements, $\lambda_{mn}(t)$ ($m\neq n$), lead to   ``direct" relaxation processes.

In what follows, we restrict ourselves to diagonal noise effects produced by two independent (uncorrelated) protein environments described by the RTPs, $\xi_{1,2}(t)$. Then, one can write, $\lambda_{mn}(t) = \delta_{mn} \sum^2_{a=1}\lambda^{(a)}_n \xi_a(t) $,  where, $\lambda^{(a)}_n$, is the interaction constant with the $a$-th environment, $\xi_a(t)$, at the site, $n$ ($a,n=1,2$). Note, that in our approach, each noise can act on both donor and acceptor sites. (See Fig. \ref{TLS}.) The limit of a single collective noise, acting on both the donor and acceptor sites, corresponds to: $\lambda_{1,2}^{(1)}\neq 0$ and $\lambda_{1,2}^{(2)}= 0$, or $\lambda_{1,2}^{(1)}= 0$ and $\lambda_{1,2}^{(2)}\neq 0$. The limit of two uncorrelated noises, acting one on the donor and another on the acceptor, corresponds to: $\lambda_{1}^{(2)}=\lambda_{2}^{(1)}=0$, or $\lambda_{1}^{(1)}=\lambda_{2}^{(2)}=0$.

We consider the stationary telegraph noise described by the random variable,  $\xi_a(t) = \zeta_a(t) -\bar{\zeta}_a $, so that,
 \begin{align}\label{chi_8a}
&\langle \xi_a(t)\rangle =0, \\
 &\langle \xi_a(t)\xi_b(t')\rangle = \delta_{ab}\chi_a(t-t'),
\end{align}
where, $\chi_a(t-t') = \sigma_a^2 e^{-2\gamma_a |t-t'|}$, is the correlation function of $a$-th noise, described by the random variable, $\xi_a(t)$.  The average value, $\langle \zeta_a(t)\rangle =\bar{\zeta}_a$, is included in the renormalization of the electron energy at each site, $n$, in Eq. (\ref{Eq2b}) as: $\varepsilon_n\rightarrow \varepsilon_n+ \sum_a \lambda^{(a)}_n \bar{\zeta}_a$.\\

\subsection{Integro-differential equations and rates }

The dynamics of the TLS can be described by the following system of  integro-differential equations \cite{14,Nes2}.  (For details see the SM.):
\begin{align} 
\frac{d}{dt}\langle{{\rho}}_{11}(t)\rangle =&- \int_0^t { K}(t-t') (\langle{\rho}_{11}(t')\rangle -\langle{\rho}_{22}(t')\rangle)dt'
\nonumber \\
&+ iV_{21}\langle\rho_{12}(0)\rangle - iV_{12}\langle\rho_{21}(0)\rangle, \label{DC4a}\\
\frac{d}{dt}\langle{{\rho}}_{22}(t)\rangle  =& \int_0^t { K}(t-t') (\langle{\rho}_{11}(t')\rangle -\langle{\rho}_{22}(t')\rangle )dt'
\nonumber \\
&- iV_{21}\langle\rho_{12}(0)\rangle + iV_{12}\langle\rho_{21}(0)\rangle,
\label{DC4b}
\end{align}
where, 
\begin{align}\label{YDAK1}
  K(t-t') = & 2V^2\Phi(t-t') \cos (\varepsilon(t-t')),
\end{align}
$\varepsilon = \varepsilon_1 - \varepsilon_2$, and $\Phi(t-t')$, is the characteristic functional of the random process.

For the case of two uncorrelated environments described by the RTPs, one can show that $\Phi(t) =\Phi_1(t)\Phi_2(t)$. The characteristic  functional, $\Phi_a(t)$, of each independent RTP, is given by \cite{BGA,NB1},
\begin{widetext}
\begin{align} \label{AP1}
\Phi_a(t) = e^{-\gamma_a t}\Big (  \cosh\big(\sqrt{\gamma^2_a- d_a^2}\,{ t}\big)  + \frac{1}{\sqrt{\gamma^2_a- d_a^2}}  \sinh\big(\sqrt{\gamma^2_a- d_a^2}\,{ t}\big) \Big ), \quad a =1,2,
\end{align}
\end{widetext}
where, $d_a = (\lambda^{(a)}_1- \lambda^{(a)}_2 )\sigma_a$, denotes the amplitude of $a$-th noise.

When the condition, $|\int_0^\infty \tau K(\tau)d\tau |\ll1$, is satisfied,  we can approximate Eqs. (\ref{DC4a}) and (\ref{DC4b}) by the following system of ordinary differential equations,
\begin{align} 
\frac{d}{dt}{\langle{\rho}}_{11}(t)\rangle =&-{R}(t)\big(\big\langle{\rho}_{11}(t)\big\rangle -\big\langle{\rho}_{22}(t)\big\rangle\big)  \nonumber \\
&+ iV_{21}\langle\rho_{12}(0)\rangle - iV_{12}\langle\rho_{21}(0)\rangle, 
\label{N5}\\
\frac{d}{dt}{\langle{\rho}}_{22}(t)\rangle =&{R}(t)\big(\big\langle{\rho}_{11}(t)\big\rangle -\big\langle{\rho}_{22}(t)\big\rangle\big)  \nonumber \\
&- iV_{21}\langle\rho_{12}(0)\rangle + iV_{12}\langle\rho_{21}(0)\rangle,
\label{N5a}
\end{align}
where ${R}(t)= \int_{0}^{t} K (\tau) d\tau$.

 Assume that initially the off-diagonal components of the density matrix (and, correspondingly, their average values) are zero, $\rho_{12}(0) = \rho_{21}(0) =0$. Then, the exact solution of Eqs.  (\ref{N5}) and (\ref{N5a}) can be written as:
\begin{align}\label{Bt}
\big\langle{\rho}_{11}(t)\big\rangle = \frac{1}{2} + \bigg(\langle\rho_{11}(0)\rangle-\frac{1}{2}\bigg)\displaystyle e^{-2\int_0^t {R}(t')dt'}, \\
\big\langle{\rho}_{22}(t)\big\rangle = \frac{1}{2} + \bigg(\langle\rho_{22}(0)\rangle -\frac{1}{2}\bigg)\displaystyle e^{-2\int_0^t  {R}(t')dt'}, \label{B8}
\end{align}
where, $\langle\rho_{11}(0)\rangle=\rho_{11}(0)$ and $\langle\rho_{22}(0)\rangle=\rho_{22}(0)$.
As one can see, in the limit $t\rightarrow\infty$, the presence of noise results in equal populations in the TLS. 

The solution given by Eqs. (\ref{Bt}) and (\ref{B8}) can be approximated by replacing ${R}(t)$ by its asymptotic value, $\Gamma/2 = \lim_{t \rightarrow \infty} {R}(t)$. The result is,
\begin{align}\label{B1}
\big\langle{\rho}_{11}(t)\big\rangle = \frac{1}{2} + \bigg(\rho_{11}(0)-\frac{1}{2}\bigg)\displaystyle e^{-\Gamma t}, 
\\
\big\langle{\rho}_{22}(t)\big\rangle = \frac{1}{2} + \bigg(\rho_{22}(0) -\frac{1}{2}\bigg)\displaystyle e^{-\Gamma t}. \label{B2}
\end{align}

{\it Two uncorrelated noises.} When the environment is described by two uncorrelated RTPs, the asymptotic rate, $\Gamma$, is given by (see the SM for details),
\begin{widetext}
\begin{align}
\Gamma=&\displaystyle\frac{2|V_{12}|^2}{\alpha _{1}\,\alpha _{2}\, (\gamma_1 + \gamma_2)}\Re\Bigg(\displaystyle \,{\frac { \left( g_{1}\,g_{2}-\alpha _{1}\,\alpha _{2} \right)  \left( 1+i\nu \right) + \left( \alpha _{1}\,g_{2} -g_{1}\,\alpha _{2}\right)  \left( \alpha _{1}-\alpha _{2} \right) }{ \left( \alpha _{1}-\alpha _{2} \right) ^{2}- \left( 1+i\nu \right) ^{2}  }} \nonumber \\
&- \displaystyle \,{\frac { \left( g_{1}\,g_{2}+\alpha _{1}\,\alpha _{2} \right)  \left( 1+i\nu \right) + \left( g_{1}\,\alpha _{2}+\alpha _{1}\,g_{2} \right)  \left( \alpha _{1}+\alpha _{2} \right) }{\left( \alpha _{1}+\alpha _{2} \right) ^{2}- \left( 1+i\nu \right) ^{2}  }}\Bigg),
\label{R3d}
\end{align}
where,
\begin{align}
&\alpha_1= \sqrt{g_1^2 -\mu_1^2}, \quad \alpha_2= \sqrt{g_2^2 -\mu_2^2}, \quad g_1= \frac{\gamma_1 }{\gamma_1 + \gamma_2}, \quad g_2= \frac{\gamma_2 }{\gamma_1 + \gamma_2}, \nonumber \\
& \displaystyle \nu=\frac{\varepsilon}{(\gamma_1 + \gamma_2)}, \quad \mu_1=\frac{d^{(1)}_1 -d^{(1)}_2}{\gamma_1 + \gamma_2}, \quad \mu_2=\frac{d^{(2)}_1 -d^{(2)}_2}{\gamma_1 + \gamma_2} .
\label{R2a}
\end{align}
\end{widetext}
In Eq. (\ref{R3d}), we used the following notation: $d^{(a)}_n = \lambda^{(a)}_n \sigma_a$ ($a,n=1,2$). 

As was mentioned above, the condition of the applicability of Eqs. (\ref{N5}) and (\ref{N5a}) is:  $|\int_0^\infty \tau K(\tau)d\tau |\ll1$. To analyze this condition analytically is rather complicated. Our numerical calculations show the approximate condition of applicability is: $|V_{1,2}|\lesssim (\gamma_1+\gamma_2)$.  Note, that the same condition is also required for the exact solutions of Eqs. (\ref{A4a})-(\ref{A4b}) to be approximated by the Eqs. (\ref{B1}) and (\ref{B2}). (See for details the SM.)

{\it Single collective diagonal noise}. In the case of a single collective noisy environment, acting on both the donor and acceptor, the rate, $\Gamma$, in Eq. (\ref{R3d}) has the form,
\begin{align}\label{R2b}
 {\Gamma}= \frac{8V^2\mu^2 }{\gamma((\mu^2-\nu^2 )^2+ 4\nu^2)}.
\end{align}
Substituting $\mu=d/\gamma$ and  $\nu=\varepsilon/\gamma$,  we obtain, 
\begin{align}\label{G1}
 {\Gamma}= \frac{8\gamma |V_{12}|^2 d^2 }{(d^2-\varepsilon^2 )^2+ 4\gamma^2\varepsilon^2},
\end{align}
where,  $d=(\lambda_1 - \lambda_ 2)\sigma$, denotes the amplitude of the noise. As one can see, the rate, $\Gamma$, reaches its  maximum, 
\begin{align}\label{max}
\Gamma_{max} =\frac{4\gamma |V_{12}|^2}{\sqrt{\varepsilon^4+4\gamma^2\varepsilon^2}- \varepsilon^2},
\end{align}
at the ``resonance" amplitude of noise, 
\begin{align}\label{res}
d_{res}=(\varepsilon^4+4\gamma^2\varepsilon^2)^{1/4}.
\end{align}
(See also \cite{6,7}.) When the amplitude of noise is far from the resonance value, the rate, $\Gamma$, becomes very small.

{\it The ``nonlinear" regime of electron transfer.} The dependence of the ET rate, $\Gamma$, in Eq. (\ref{G1})  on the amplitude, $d$, of noise (the external random force), is a nonlinear one.  Indeed, the amplitude, $d$, appears in $\Gamma$ in both the numerator and in the denominator. Suppose, that the value of $d$ is small ($d\ll\varepsilon$). In this case, $\Gamma\approx 8\gamma|V_{12}|^2/\varepsilon^2(\varepsilon^2+4\gamma^2)$, and the rate is proportional to the intensity of the external random process. So, in this ``linear" regime (small $d$), there are no resonances in the $\Gamma(d)$ behavior. In the opposite case of  strong noise ($d\gg\varepsilon$), the ET rate is: $\Gamma\approx 8\gamma|V_{12}|^2/d^2$, and it decreases as $d$ increases. We can say that the strong noise does not allow the electron to move from the donor to the acceptor, a kind of ET Zeno effect. Only for the intermediate noise amplitudes, $d$, the ``resonance" in the behavior of $\Gamma(d)$ takes place. In this sense, the regime of the ET is a nonlinear one. A similar situation occurs when two uncorrelated noises are applied to the system. In this case, two ``interacting nonlinear resonances" occur.

There are two limiting cases in which the expression for $\Gamma_{max}$ can be simplified. (1) $\varepsilon\gg 2\gamma$. In this case, $\Gamma_{max}\approx 2|V_{12}|^2/\gamma$. (2) $\varepsilon\ll 2\gamma$. In this case, $\Gamma_{max}\approx 2|V_{12}|^2/\varepsilon$. An approximate condition for applicability of Eq. (\ref{G1}) for the rate, is: $|V_{12}|\lesssim \gamma, d$.

\begin{figure}[tbh]
	\scalebox{0.3}{\includegraphics{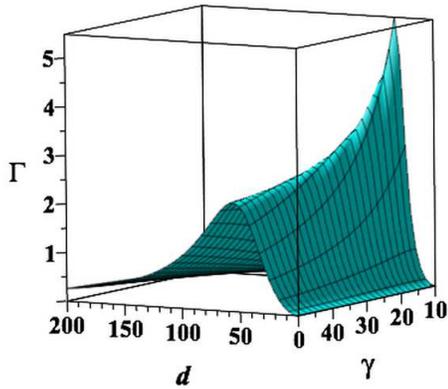}}
	\caption{(Color online) Dependence of the rate, $\Gamma$, in Eq. (\ref{G1}) on noise amplitude, $d$, and the correlation rate, $\gamma$.  Parameters: $V_{12}=5$, $\varepsilon= 30$.
		\label{R3}}
\end{figure}

\begin{figure}[tbh]
	\scalebox{0.275}{\includegraphics{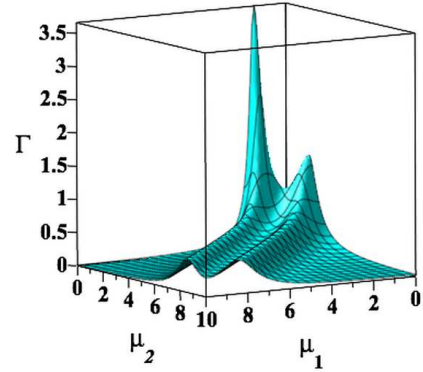}}
	\scalebox{0.275}{\includegraphics{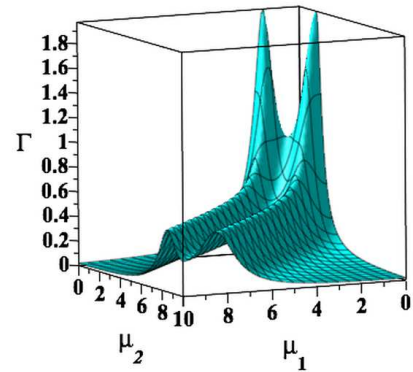}}
	\caption{(Color online) Asymptotic rate, $\Gamma$, of Eq. (\ref{R3d}) vs. the dimensionless amplitudes of two uncorrelated noises, $\mu_1$,  and $\mu_2$.  Parameters: $V_{12}=3$, $\varepsilon= 30$. Top: $\gamma_1=5$, $\gamma_2=15$. Bottom: $\gamma_1=\gamma_2=10$.
		\label{R1}}
\end{figure}

\begin{figure}[tbh]
	\begin{center}
		\scalebox{0.325}{\includegraphics{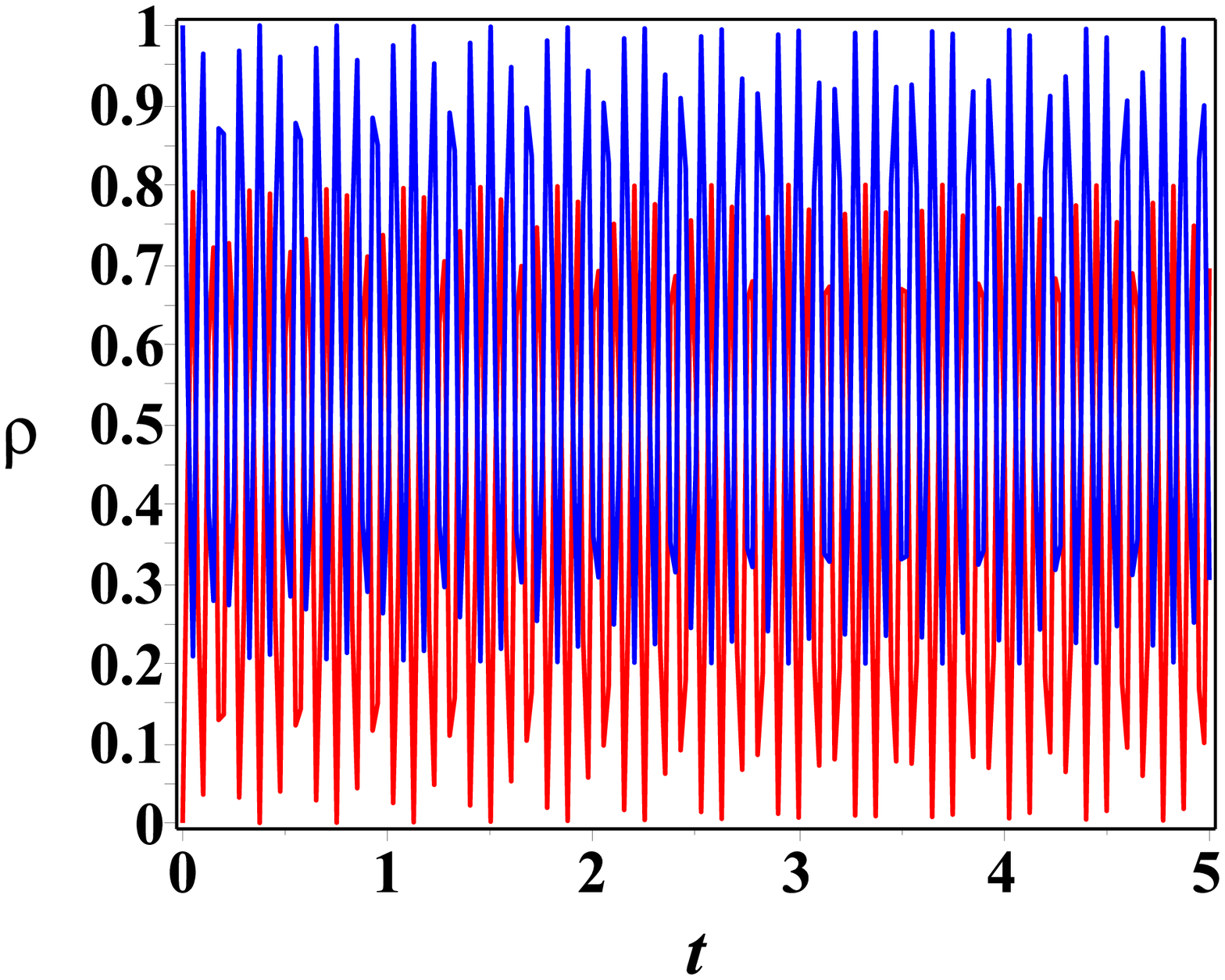}}
		(a)
		\scalebox{0.325}{\includegraphics{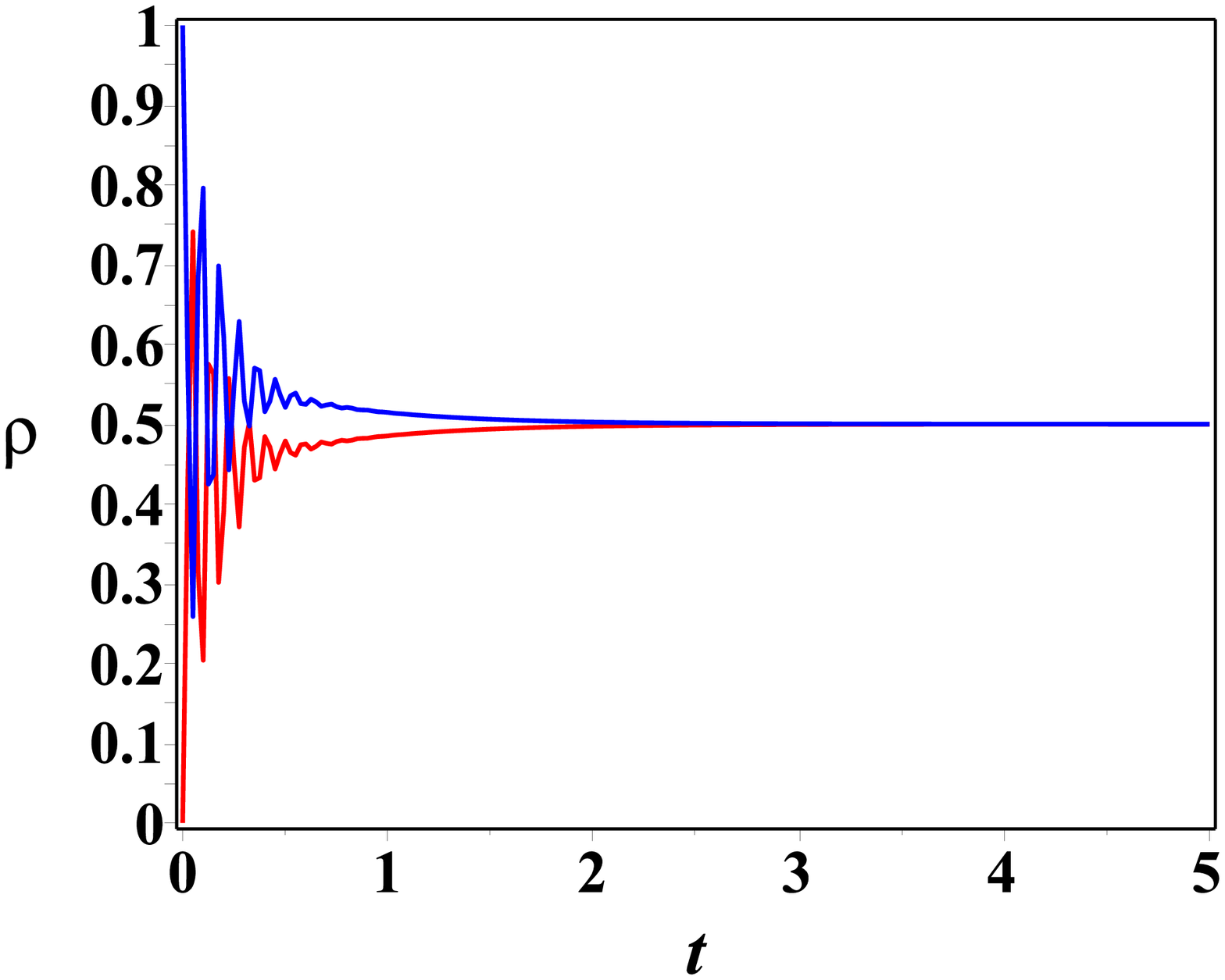}}
		(b)
	\end{center}
	\caption{(Color online) Strongly coupled dimer ($\mu_d=1$). Time dependence (in ps) of the density matrix components:  $\rho_{11}(t)$ (blue) and $\rho_{22}(t)$ (red).  Parameters: $V_{12}=30$, $\varepsilon_1= 60, \varepsilon_2 = 30$,  $\gamma_1=10$, $\gamma_2=15$. (a) $d^{(1)}_1= 10$, $d^{(1)}_2= 10$,  $d^{(2)}_1= 0$, $d^{(2)}_2= 0$, (b) $d^{(1)}_1= 10$, $d^{(1)}_2= 0$,  $d^{(2)}_1= 0$, $d^{(2)}_2= 10$. Initial conditions:  $\rho_{11}(0) =1$,  $\rho_{22}(0) =0$.
		\label{F1}}
\end{figure}

\begin{figure}[tbh]
	\begin{center}
		\scalebox{0.325}{\includegraphics{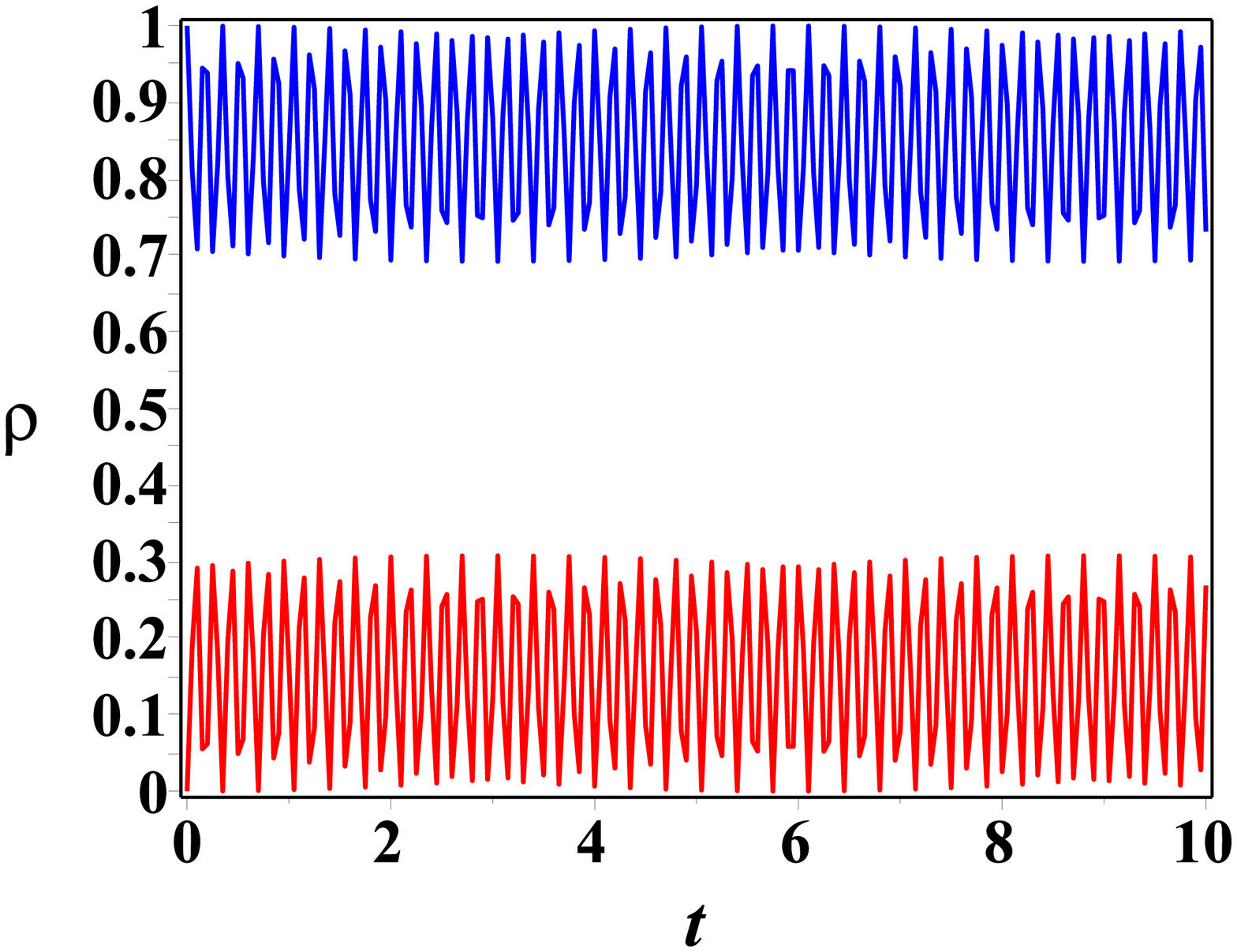}}
		(a)
		\scalebox{0.325}{\includegraphics{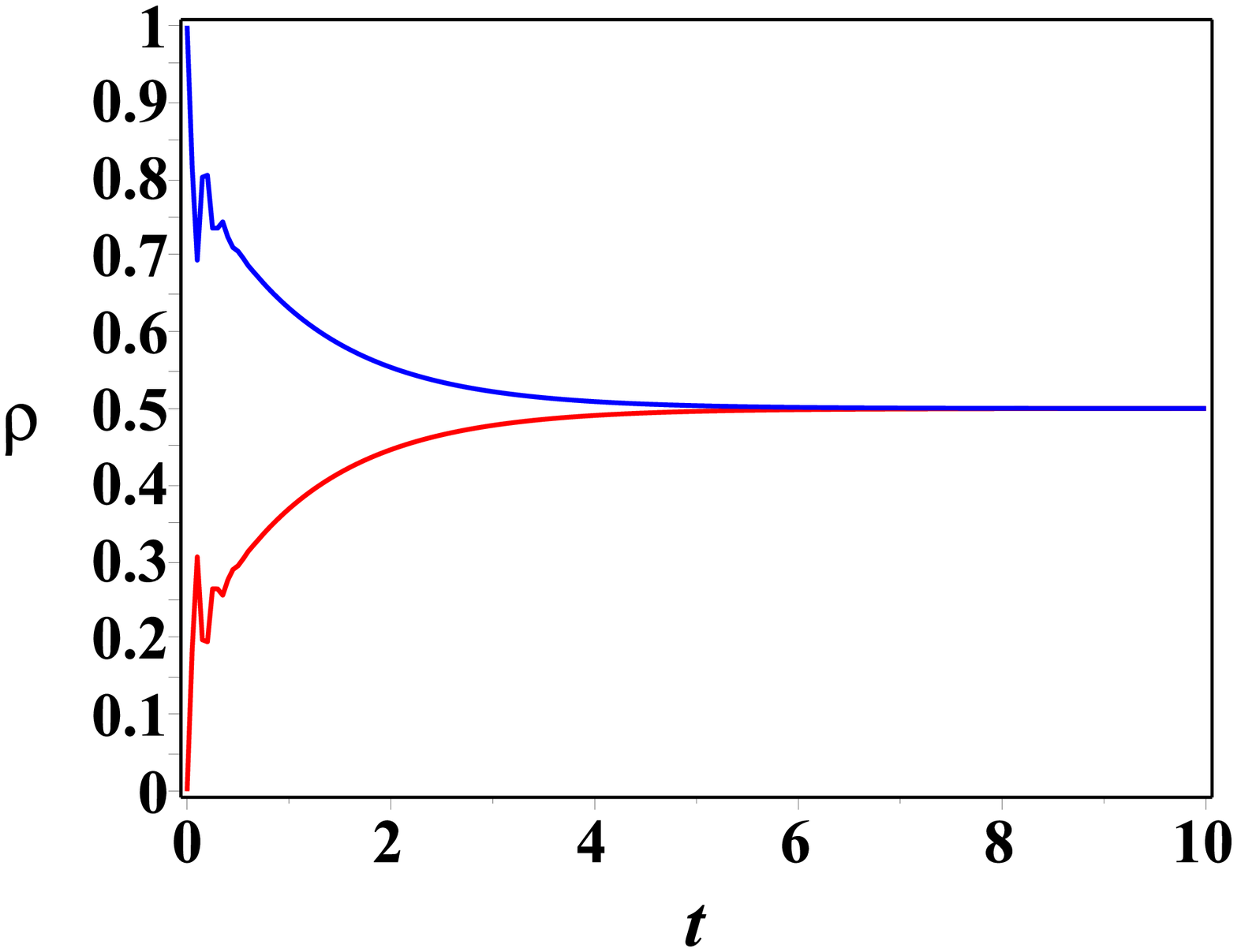}}
		(b)
	\end{center}
	\caption{(Color online) Time dependence (in ps) of the density matrix components:  $\rho_{11}(t)$ (blue) and $\rho_{22}(t)$ (red).  Parameters: $V_{12}=10$, $\varepsilon_1= 60, \varepsilon_2 = 30$,  $\gamma_1=10$, $\gamma_2=15$. (a) $d^{(1)}_1= 10$, $d^{(1)}_2= 10$,  $d^{(2)}_1= 0$, $d^{(2)}_2= 0$, (b) $d^{(1)}_1= 10$, $d^{(1)}_2= 0$,  $d^{(2)}_1= 0$, $d^{(2)}_2= 10$. Initial conditions:  $\rho_{11}(0) =1$,  $\rho_{22}(0) =\rho_{12}(0)=0$.
		\label{F2}}
\end{figure}

\subsection{Results of numerical simulations}
In the numerical simulations, it is convenient to measure the energy parameters in units of $\rm ps^{-1}$, while time is measured in $\rm ps$. Then, the energy $\varepsilon = 1\rm ps^{-1}\approx 0.66\rm meV$.

In Fig. \ref{R3}, we show the rate, $\Gamma$, defined by Eq. (\ref{G1}) (for a single noise), as a function of the amplitude of noise, $d$, and the correlation rate of noise (inverse correlation time), $\gamma$. As one can see, $\Gamma$ reaches its maximum value at the resonance amplitude of noise, $d_{res}$, given by Eq. (\ref{res}). At the same time, as one can see from Eq. (\ref{res}), for a given value of the redox potential, $\varepsilon$,  the value of $\Gamma_{max}$ depends of $\gamma$. This behavior is demonstrated in Fig. \ref{R3}, for $\varepsilon=50$, $|V_{1,2}|=5$, and for $10\leqslant\gamma\leqslant 50$. One case see, that for these parameters, $\Gamma_{max}\lesssim 6\rm ps^{-1}$. 

\begin{figure}[tbh]
	\begin{center}
		\scalebox{0.325}{\includegraphics{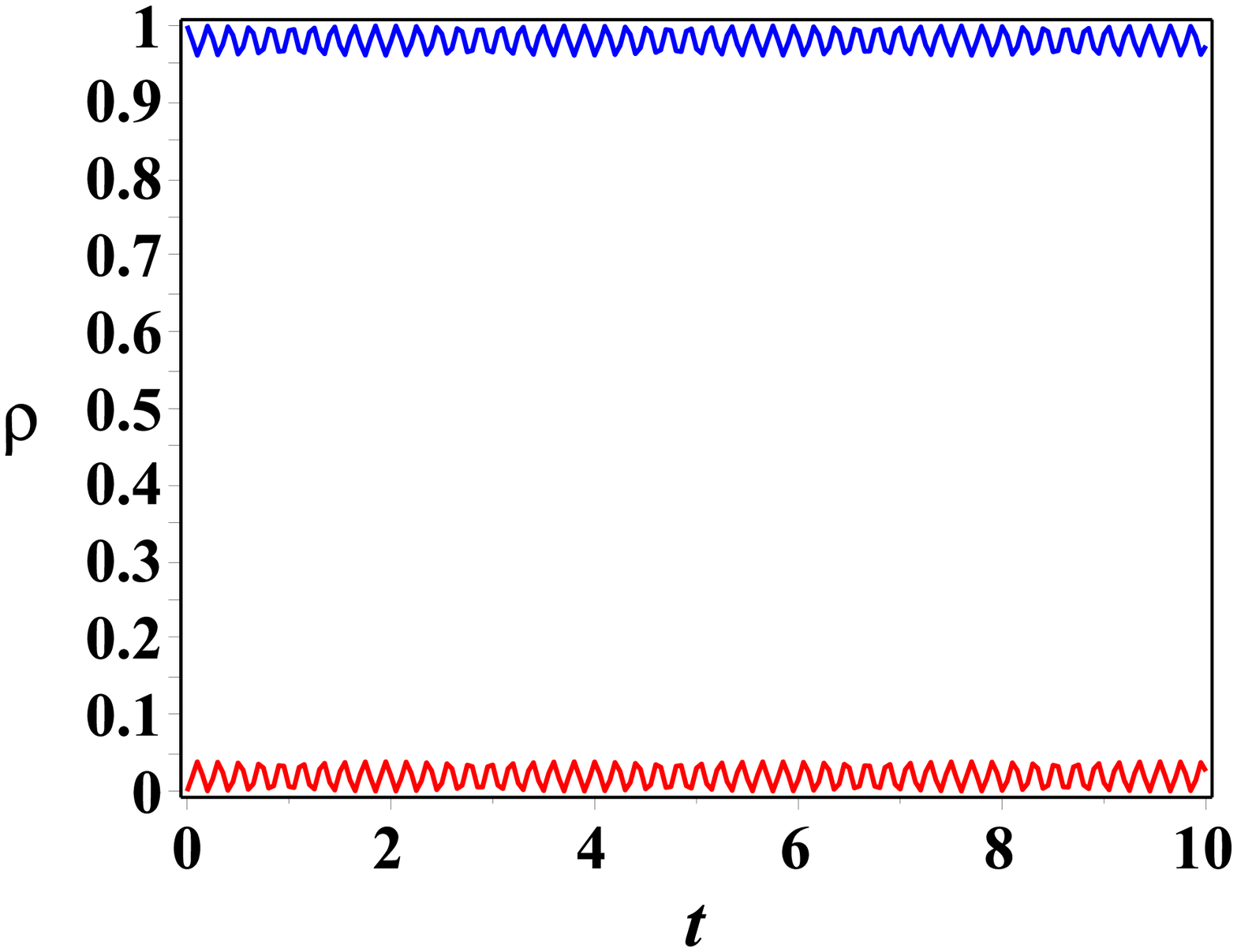}}
		(a)
		\scalebox{0.325}{\includegraphics{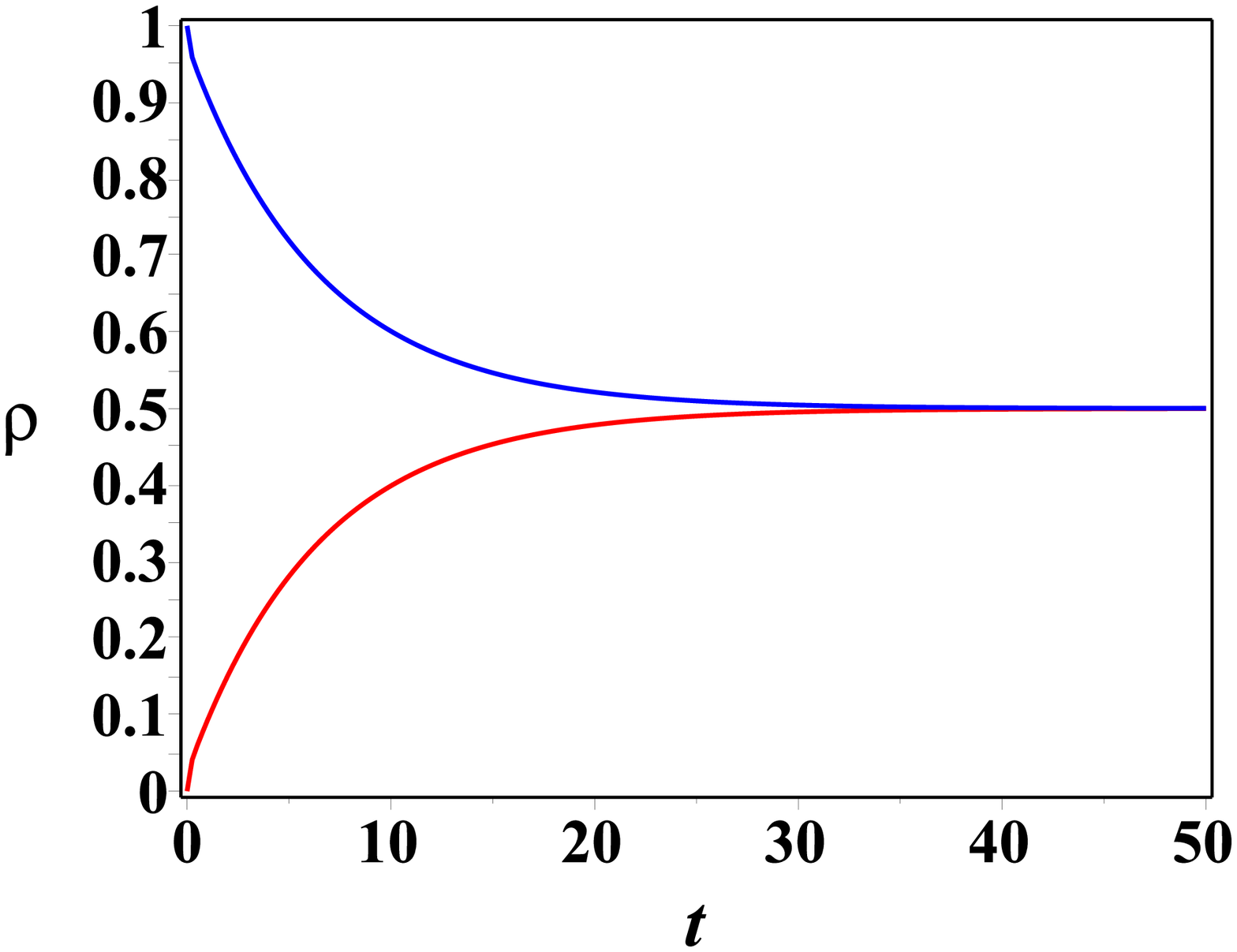}}
		(b)
	\end{center}
	\caption{(Color online) Weakly coupled dimer ($\mu_d=0.1$). Time dependence (in ps) of the density matrix components:  $\rho_{11}(t)$ (blue) and $\rho_{22}(t)$ (red).  Parameters: $V_{12}=3$, $\varepsilon_1= 60, \varepsilon_2 = 30$,  $\gamma_1=10$, $\gamma_2=15$. (a) $d^{(1)}_1= 10$, $d^{(1)}_2= 10$,  $d^{(2)}_1= 0$, $d^{(2)}_2= 0$, (b) $d^{(1)}_1= 10$, $d^{(1)}_2= 0$,  $d^{(2)}_1= 0$, $d^{(2)}_2= 10$. Initial conditions:  $\rho_{11}(0) =1$,  $\rho_{22}(0) =rho_{12}(00=0$.
		\label{F3}}
\end{figure}

\begin{figure}[tbh]
	\begin{center}
		\scalebox{0.35}{\includegraphics{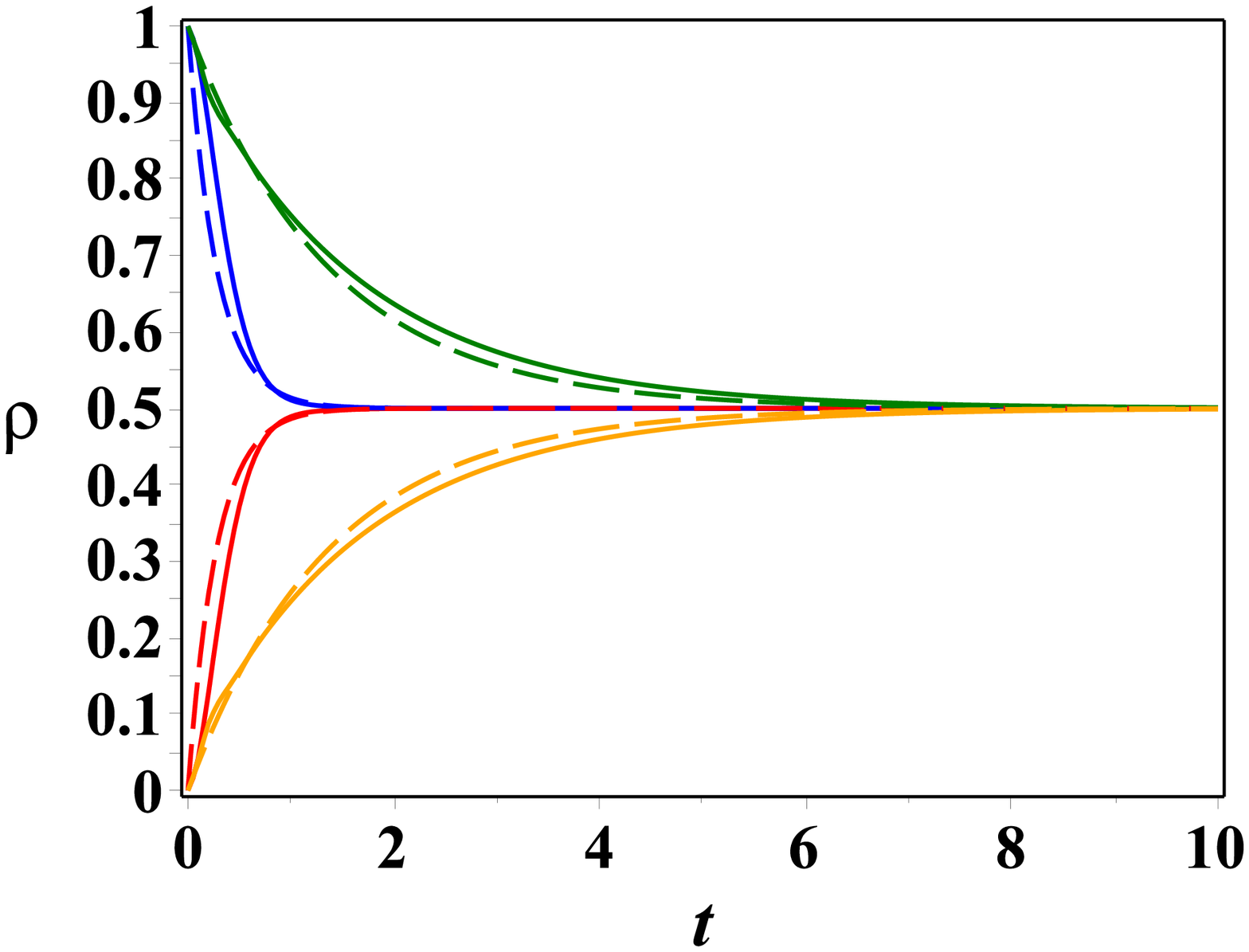}}
	\end{center}
	\caption{(Color online) Weakly coupled dimer ($\mu_d=0.1$). Time dependence (in ps) of the density matrix components:  $\rho_{11}(t)$ (blue and green curves), $\rho_{22}(t)$ (red and orange curves).  Choice of parameters: $V_{12}=3$, $\varepsilon_1= 60, \varepsilon_2 = 30$,  $\gamma_1=5$, $\gamma_2=15$. Blue and red curves: $d^{(1)}_1= 20$, $d^{(1)}_2= -10$,  $d^{(2)}_1= 0$, $d^{(2)}_2= 0$, $\Gamma= 3.6$, $(\mu_1=1.5;\mu_2=0)$.  Green and orange curves: $d^{(1)}_1= 20$, $d^{(1)}_2= 0$,  $d^{(2)}_1= 0$, $d^{(2)}_2= -10$,  $\Gamma= 0.73$, $(\mu_1=1;\mu_2=0.5)$.  Solid curves correspond to the solutions of the exact Eqs. (\ref{A4a})-(\ref{A4b}). Dashed curves correspond to the approximate solutions given by Eqs. (\ref{B1}) and (\ref{B2}). Initial conditions:  $\rho_{11}(0) =1$,  $\rho_{22}(0) =\rho_{12}(0)=0$.
		\label{F4}}
\end{figure}

For two uncorrelated noises, the rate, $\Gamma$, in Eq. (\ref{R3d}), is shown in Fig. \ref{R1}, as a function of two dimensionless amplitudes of noise, $\mu_1$ and $\mu_2$. As one can see,  two ``interacting" resonances are present. The amplitudes of these resonances  depend on the values of $\mu_{1,2}$. These resonances can be either non-symmetric, as in Fig. \ref{R1} (top), for different correlation rates, $\gamma_1=5$ and   $\gamma_2=15$, or symmetric, as in Fig. \ref{R1} (bottom), for equal correlation rates, $\gamma_{1,2}=10$. Note, that in both cases,  $\gamma_1+\gamma_2=20$. For these chosen parameters, $\Gamma_{max}\lesssim 3.6\rm ps^{-1}$. 

{\it Weakly and strongly coupled dimers.} The ``donor-acceptor" system shown in Fig. \ref{TLS}, represents a coupled dimer (realized, for example, by two coupled chlorophyll molecules.) This dimer can be either weakly or strongly coupled. Let us introduce the parameter, $\mu_d=|V_{12}/\varepsilon|$. It is easy to see that when $\mu_d\ll 1$, both eigenstates, $|u_{+}\rangle$ and  $|u_{-}\rangle$, of the Hamiltonian, ${\mathcal H}_0$, in (\ref{Meq2a}),  become close to the unperturbed states, $|u_{1}\rangle$ and  $|u_{2}\rangle$, when $V_{12}=0$. In this case, we call the dimer ``weakly coupled". The dimer is called ``strongly coupled", when  the value of $\mu_d$ is not too small. We can say that the dimer is strongly coupled when, $\mu_d\gtrsim 1$.

In Figs. \ref{F1} - \ref{F4}, we present the results of the numerical simulations of the dynamical behavior of the system shown in Fig. \ref{TLS}, for different parameters, and for both correlated and uncorrelated noisy environments. All simulations were performed using the exact system of equations (\ref{A4a}) - (\ref{A4b}). In Fig. \ref{F4}, we also compare the exact results with the corresponding approximate solutions. (In the SM, more details on the comparison of the exact and approximate solutions are presented.) We also consider  weakly and strongly coupled dimers. For simplicity, in all cases, the initial conditions were chosen when the donor was populated: $\rho_{11}(0) =1$,  $\rho_{22}(0) =\rho_{12}(0)=0$.

In Fig. \ref{F1}a, a single correlated noise, corresponding to $a=1$, is applied to a strongly coupled dimer $(\mu_d=1$). In this case, both amplitudes of noise, acting on donor and acceptor,  are equal, $d^{(1)}_1=d^{(1)}_2=10$. So, the effective noise,  acting on the system, is absent: $d^{(1)}=d^{(1)}_1 -d^{(1)}_2=0$. This regime is easy to understand, as the matrix, $\mathcal V_a$, in Eq. (\ref{Neq1}) becomes the unit matrix. In this case, the dynamics of the system exhibits Rabi oscillations. Because the matrix element is equal to the redox potential, $V_{1,2}=\varepsilon=30$,  the Rabi oscillations have large amplitude. The situation changes significantly when two uncorrelated noises, with the same amplitudes, $d^{(1)}_1=10$ (applied to the donor) and $d^{(2)}_2=10$ (applied to the acceptor), influence the same dimer. (See Fig. \ref{F1}b.) In this case, the dynamics experiences rapid relaxation, and saturates ($\rho_{1,2}(t)\rightarrow 1/2$) at approximately, $t_s\approx 2\rm ps$. We can conclude, that in this case, two uncorrelated noises (environments), with equal amplitudes, are more effective in assisting the ET than a single correlated noise with the same amplitude. Similar results are shown in Fig. \ref{F2}, for the intermediately coupled dimer ($\mu_d=1/3$). In this case, the amplitude of the Rabi oscillations in Fig. \ref{F2}a decreases, and the saturation time in Fig. \ref{F2}b increases, $t_s\approx 6\rm ps$. In both cases, shown in Fig. \ref{F1}b and Fig. \ref{F2}b, the electron transfer dynamics is accompanied by coherent oscillations of the populations, $\rho_{11}(t)$ and $\rho_{22}(t)$. In Fig. \ref{F3}, the case of a weakly coupled dimer is demonstrated, for $\mu_d=0.1$, and for the same amplitudes of the noisy environments as in Figs. \ref{F1} and \ref{F2}. As one can see, the amplitude of the Rabi oscillations in Fig. \ref{F3}a decreases significantly (less than $0.05$), and the saturation time  of the ET in Fig. \ref{F3}b increases significantly,  $t_s\approx 35\rm ps$. The populations, $\rho_{11}(t)$ and $\rho_{22}(t)$, do not experience visible oscillations in this case.

In Fig. \ref{F4}, we show the dynamics  of the ET for two noises, which act  on both donor and acceptor, but with different amplitudes and interaction constants. Solid curves correspond to the solutions of the exact Eqs. (\ref{A4a})-(\ref{A4b}). Dashed curves correspond to the approximate solutions given by Eqs. (\ref{B1}) and (\ref{B2}). Initial conditions are the same for all cases presented in  Fig. \ref{F4}: $\rho_{11}(0) =1$,  $\rho_{22}(0) = \rho_{12}(0)=0$. Blue and red curves correspond to the noise amplitudes: $d^{(1)}_1= 20$, $d^{(1)}_2= -10$,  $d^{(2)}_1= 0$, $d^{(2)}_2= 0$. In this case, in even though two noises are present in the system, the second noise (with $a=2$) has zero constants of interaction with both the donor and acceptor: $\lambda^{(2)}_{1,2}=0$. So, effectively only one collective noise (with $a=1$) acts on both the donor and acceptor. The dimensionless  amplitudes of noise are: $(\mu_1=1.5;\mu_2=0)$, and the ET rate is: $\Gamma= 3.6$. 

Green and orange curves correspond to the amplitudes of noise: $d^{(1)}_1= 20$, $d^{(1)}_2= 0$,  $d^{(2)}_1= 0$, $d^{(2)}_2= -10$. In these case, both noises act on the system.  The dimensionless amplitudes of noise are: $(\mu_1=1;\mu_2=0.5)$, and the ET rate is: $\Gamma=0.73$. The presented ET rates correspond to the results shown in Fig. (\ref{R1}). As one case see, for chosen parameters, the results of the approximate solutions (dashed curves) are in good agreement with the results of exact equations  (solid curves). As our results demonstrate, the saturation time depends significantly on (i) the presence of collective or independent (uncorrelated) noises acting on the donor and acceptor, (ii) the amplitude of noises, and (iii) the interaction constants with noises. Indeed, for parameters chosen in  Fig. \ref{F4}, for blue and red curves, the saturation time is: $t_{sat}\approx 1.8\rm ps$. For parameters chosen for green and orange curves, the saturation time is: $t_{sat}\approx 8\rm ps$.

\section{Conclusion}

   When modeling the  primary quantum exciton transfer processes in photosynthetic complexes, two major problems occur. The first is related to strong pigments-protein interactions. The second problem is related to the large number of pigments (or light-sensitive sites) in the light-harvesting complexes. This results in a multi-scale electron transfer dynamics and in the necessity to develop adequate coarse-grained procedures. Moreover, the number, $N$, of light-sensitive pigments in the sub-complexes of plants and algae  is neither small or large, but rather of an intermediate value, $N\approx 10-20$  \cite{1,2}. Then, it is difficult to apply the well-developed methods from the solid state physics which are used either for rare impurities or for systems with electron band structures. In this situation, it is useful to design an exactly solvable (at least, numerically) quantum model which can be used to describe the electron transfer in these complex biological systems. Such a model is introduced in this paper, for rather general photosynthetic complexes, and for any values of the ``pigment-protein" interaction constants. We considered both regimes of correlated and uncorrelated  random protein fluctuations acting on different pigments (electron sites). Both of these regimes can be realized in real photosynthetic organisms. We demonstrated that the uncorrelated protein fluctuations can either increase or decrease the electron transfer rates.  They can also modify the whole electron transfer dynamics. We also derived analytical expressions for the ET rates and for the evolution of the density matrix, which approximate the exact solutions for large time-intervals for a wide range of parameters. 

   Our model can easily be applied for many concrete light-harvesting complexes and reaction centers. The solutions which follow from our model can be used for developing adequate coarse-grained procedures, and for comparison with the results of different approximations and perturbation approaches. Our results can also be used for engineering the protein environment to achieve desired properties for the ET dynamics. In order to verify the properties of the protein environment, standard molecular dynamics (MD) methods can be used to simulate the time-dependent correlation functions between different electron sites. The generalization of our approach for thermal protein environments is one focus of our future research. One way to do this, is to develop a perturbation theory not by the constants of interactions between the electron sites and the protein fluctuations, but by the matrix elements of the interactions between different electron sites. This research is now in progress. \\
   
  \acknowledgements
  
   This work was carried out under the  auspices of the National Nuclear Security Administration of the U.S. Department of Energy at Los Alamos National Laboratory under Contract No. DE-AC52-06NA25396.
 A.I.N. acknowledges the support from the CONACyT, Grant No. 15349.

\appendix

\begin{widetext}

\section{ SUPPLEMENTARY MATERIAL}

\subsection*{Two-level system with a single noisy environment}

Here, we present mathematical details of our approach. We start with the simplified model  of two-level system (TLS) with a stationary diagonal noise, described by a random  variable,  $\xi(t)$.  The Hamiltonian of the TLS is,
\begin{align}\label{C1}
 {\mathcal H}= &(E_1+\lambda_1 \xi(t))|1\rangle\langle 1|  +  (E_2 +\lambda_2 \xi(t))|2\rangle\langle 2|  +  V_{12}|1\rangle\langle 2|+ V_{21}|2\rangle\langle 1|,
 \end{align}
 where, $\lambda_{1,2}$, are the interaction constants.
 Using (\ref{C1}), we obtain the following equations for the density matrix elements ($\hbar=1$),
\begin{align} \label{AX4}
{\dot \rho}_{11}(t) =&iV_{21}{\rho}_{12}(t) - iV_{12}{\rho}_{21}(t)), \\
{\dot \rho}_{22} (t)= & iV_{12}{\rho}_{21}(t))   -iV_{21}{\rho}_{12}(t) ,\label{AX4c}\\
{\dot \rho}_{12}(t)  = &- i(\varepsilon  + D\xi(t)){\rho}_{12}(t)  +iV_{12}( {\rho}_{11}(t)-{\rho}_{22}(t) ), \label{AX4b}\\
{\dot \rho}_{21}(t)  = & i(\varepsilon  + D\xi(t)){\rho}_{21}(t)  -iV_{21}( {\rho}_{11}(t)-{\rho}_{22}(t) ),
\label{AX4a}
\end{align}
where $\varepsilon =E_1 -E_2$ and $D=\lambda_1- \lambda_2$. 

After averaging over the random process, we obtain,
\begin{align} \label{AX5}
\frac{d}{dt}{\langle{\rho}}_{11}(t)\rangle =&iV_{21}\langle{\rho}_{12}(t)\rangle - iV_{12}\langle{\rho}_{21}(t)\rangle , \\
\frac{d}{dt}{\langle{\rho}}_{22}(t)\rangle =&iV_{12}\langle{\rho}_{21}(t)\rangle -iV_{21}\langle{\rho}_{12}(t)\rangle  ,\\
\frac{d}{dt}{\langle{\rho}}_{12}(t)\rangle = &- i\varepsilon \langle{\rho}_{12}(t)\rangle - i D\langle \xi(t){\rho}_{12}(t)\rangle  +iV_{12}(\langle {\rho}_{11}(t)\rangle-{\langle\rho}_{22}(t)\rangle ), \label{AX5b}\\
\frac{d}{dt}{\langle{\rho}}_{21}(t)\rangle  = & i\varepsilon \langle{\rho}_{21}(t)\rangle + iD\langle \xi(t){\rho}_{21}(t)\rangle  -iV_{21}( \langle{\rho}_{11}(t)\rangle - \langle{\rho}_{22}(t)\rangle ),
\label{AX5a}
\end{align}
where, $\langle ...\rangle$, denotes averaging over the random process. Due to the terms,  $\langle\xi(t){\rho}_{12}(t)\rangle$ and $\langle \xi(t){\rho}_{21}(t)\rangle$, on the  RHS of Eqs. (\ref{AX5b}) and (\ref{AX5a}),   this system of differential equations is not closed.

When the noise is described by a random telegraph process (RTP), one can obtain a closed system of differential equations for averaged variables \cite{KV2,NB1}. Let $\chi(\tau) = \sigma^2 e^{-2\gamma \tau}$ be a correlation function of the RTP. Employing the differential formula \cite{KV2},
\begin{align}
\Big(\frac{d}{dt} +2\gamma\Big)\langle  {\xi_a}(t)R[t;\xi(\tau) ]
=\Big\langle  {\xi}(t)\frac{d}{dt}R[t;\xi(\tau)]
\Big\rangle ,
\end{align}
where, $R[t;\xi(\tau)]$, is an arbitrary functional, and after some algebra,  we obtain the following closed system of differential equations:
\begin{align} \label{AX6}
\frac{d}{dt}{\langle{\rho}}_{11}(t)\rangle =&iV_{21}\langle{\rho}_{12}(t)\rangle - iV_{12}\langle{\rho}_{21}(t)\rangle, \\
\frac{d}{dt}{\langle{\rho}}_{22}(t)\rangle =&iV_{12}\langle{\rho}_{21}(t)\rangle -iV_{21}\langle{\rho}_{12}(t)\rangle,\\
\frac{d}{dt}{\langle{\rho}}_{12}(t)\rangle = &- i\varepsilon \langle{\rho}_{12}(t)\rangle - i D\langle {\rho}^\xi_{12}(t)\rangle  +iV_{12}( \langle{\rho}_{11}(t)\rangle-\langle{\rho}_{22}(t)\rangle), \label{AX6a}\\
\frac{d}{dt}{\langle{\rho}}_{21}(t)\rangle  = & i\varepsilon \langle{\rho}_{21}(t)\rangle + iD\langle {\rho}^\xi_{21}(t)\rangle  -iV_{21}( \langle{\rho}_{11}(t)\rangle - \langle{\rho}_{22}(t)\rangle),
\label{AXh}\\
\frac{d}{dt}{\langle{\rho}}^\xi_{11}(t)\rangle =&iV_{21}\langle{\rho}^\xi_{12}(t)\rangle - iV_{12}\langle{\rho}^\xi_{21}(t)\rangle -2\gamma{\langle{\rho}}^\xi_{11}(t)\rangle, \label{AX6b}
\\
\frac{d}{dt}{\langle{\rho}}^\xi_{22}(t)\rangle =&  iV_{12}\langle{\rho}^\xi_{21}(t)\rangle - iV_{21}\langle{\rho}^\xi_{12}(t)\rangle- 2\gamma{\langle{\rho}}^\xi_{22}(t)\rangle,\label{AX6c}
\\
\frac{d}{dt}{\langle{\rho}}^\xi_{12}(t)\rangle = &- i\varepsilon \langle{\rho}^\xi_{12}(t)\rangle - i D\sigma^2\langle{\rho}_{12}(t)\rangle  +iV_{12}(\langle {\rho}^\xi_{11}(t)\rangle-\langle{\rho}^\xi_{22}(t)\rangle)  - 2\gamma {\langle{\rho}}^\xi_{12}(t)\rangle, \label{AX6d}\\
\frac{d}{dt}{\langle{\rho}}^\xi_{21}(t)\rangle  = & i\varepsilon \langle{\rho}^\xi_{21}(t)\rangle + iD\sigma^2\langle {\rho}_{21}(t)\rangle  -iV_{21}( \langle{\rho}^\xi_{11}(t)\rangle - \langle{\rho}^\xi_{22}(t)\rangle ) - 2\gamma{\langle{\rho}}^\xi_{21}(t)\rangle,
\label{AX6e}
\end{align}
where $\langle\rho^\xi_{ij}(t) \rangle= \langle\xi(t)\rho_{ij}(t)\rangle$.

\subsection*{Integro-differential equations and rates}

Here we obtain an approximate system of integro-differential equations for the averaged components of the density matrix.  By integrating Eqs. (\ref{AX4b}) and (\ref{AX4a}), and averaging over the random process, we obtain,
\begin{align}
\langle{\rho}_{12}(t)\rangle = &iV_{12}\int_0^t \langle e^{i\varphi(t)} e^{-i\varphi(t')} e^{-i\varepsilon (t-t')}\big(\rho_{11}(t') - \rho_{22}(t')\big)\rangle dt' + {\langle\rho}_{12}(0)\rangle, \label{T2a}\\
\langle{\rho}_{21}(t)\rangle = &-iV_{21}\int_0^t \langle e^{-i\varphi(t)} e^{i\varphi(t')} e^{-i\varepsilon (t-t')}\big (\rho_{11}(t') - \rho_{22}(t')\big)\rangle dt' + \langle{\rho}_{21}(0)\rangle, 
\label{T2b}
\end{align}
where $\varphi(t) =D\int^t_0\xi(t') dt'$.

To proceed further, one should split the correlations on the RHS of Eqs. (\ref{T2a}) and (\ref{T2b}). Let, $F(t;\xi(\tau))$, be an arbitrary functional of the random process, $\xi(\tau)$. We assume that the following relation for splitting of correlations can be used:
\begin{align}\label{SP1}
  \big \langle e^{i\varphi(t)} e^{-i\varphi(t')}F(t';\xi(\tau))\big\rangle  \approx
\langle e^{i\varphi(t)} e^{-i\varphi(t')} \big\rangle   \big \langle F(t';\xi(\tau))\big\rangle=\Phi(t-t')    \big \langle F(t';\xi(\tau))\big\rangle,
\end{align}
where, $\Phi(t-t') = \langle e^{i\varphi(t)} e^{-i\varphi(t')} \big\rangle $, is the characteristic functional of the random process. 

After splitting of correlations in Eqs. (\ref{T2a}) - (\ref{T2b}), we obtain,
\begin{align}
\langle{\rho}_{12}(t)\rangle =& iV_{12}\int_0^t \Phi(t-t') e^{-i\varepsilon (t-t')}\big (\langle\rho_{11}(t')\rangle  - \langle \rho_{22}(t')\rangle \big)dt' + \langle{\rho}_{12}(0)\rangle,
 \label{NAT2b}\\
\langle{\rho}_{21}(t)\rangle =& -iV_{21}\int_0^t \Phi(t-t') e^{i\varepsilon (t-t')}\big (\langle\rho_{11}(t')\rangle  - \langle \rho_{22}(t')\rangle \big) dt' +\langle{\rho}_{21}(0)\rangle.
\label{NAT2c} 
\end{align}
Substituting these results into Eqs. (\ref{AX5}) - (\ref{AX5a}), we obtain the following system of integro-differential equations:
\begin{align} 
\frac{d}{dt}{\langle{\rho}}_{11}(t)\rangle =&- \int_0^t { K}(t-t') (\langle{\rho}_{11}(t')\rangle - \langle{\rho}_{22}(t') \rangle )dt' + iV_{21}\langle\rho_{12}(0)\rangle - iV_{12}\langle\rho_{21}(0)\rangle, \label{DC5a}\\
\frac{d}{dt}{\langle{\rho}}_{22}(t)\rangle  =& \int_0^t { K}(t-t') (\langle{\rho}_{11}(t') \rangle- \langle{\rho}_{22}(t')\rangle)dt' - iV_{21}\langle\rho_{12}(0)\rangle + iV_{12}\langle\rho_{21}(0)\rangle,
\label{DC5b}
\end{align}
where, 
\begin{align}\label{YDAK2}
  K(t-t') =  2|V_{12}|^2\Phi(t-t') \cos (\varepsilon(t-t')).
\end{align}

When the condition, $|\int_0^\infty \tau K(\tau)d\tau |\ll1$, is satisfied,  we can approximate Eqs. (\ref{DC4a}) - (\ref{DC4b}) by the following system of ordinary differential equations,
\begin{align} \label{AD4a}
\frac{d}{dt}{\langle{\rho}}_{11}(t) \rangle=&- { R}(t)  (\langle{\rho}_{11}(t)\rangle - \langle{\rho}_{22}(t) \rangle ) + iV_{21}\langle\rho_{12}(0)\rangle - iV_{12}\langle\rho_{21}(0)\rangle,\\
\frac{d}{dt}{\langle{\rho}}_{22}(t)\rangle  =& { R}(t) ( \langle{\rho}_{11}(t)\rangle - \langle{\rho}_{22}(t)\rangle  ) - iV_{21}\langle\rho_{12}(0)\rangle + iV_{12}\langle\rho_{21}(0)\rangle,
\label{AD4b}
\end{align}
where, ${ R}(t) =\int_0^t  K(\tau)d\tau$. By excluding, $\langle\rho_{22}(t)\rangle$, we obtain,
\begin{align} \label{AS1}
\frac{d}{dt}{\langle{\rho}}_{11}(t)\rangle =&- 2{ R}(t)\langle{\rho}_{11}(t)\rangle  +{R}(t) + iV_{21}\angle\rho_{12}(0)\rangle - iV_{12}\langle\rho_{21}(0)\rangle.
\end{align}

The solution of this equation, with the initial conditions, $\rho_{11}(0)=\langle\rho_{11}(0)\rangle=1$, $\rho_{12}(0)=\langle\rho_{12}(0)\rangle=0$, is, 
\begin{align}\label{DC6b}
{\langle{\rho}}_{11}(t)\rangle = \frac{1}{2} +   \frac{1}{2} e^{-2\int^t_0{ R}(\tau)d\tau}.\end{align}
Asymptotically, as $t \rightarrow \infty$, we obtain,
\begin{align}\label{DC6ab}
{\langle{\rho}}_{11}(t)\rangle = \frac{1}{2} +   \frac{1}{2} e^{-\Gamma t},
\end{align}
where $\Gamma =2 \lim_{t\rightarrow \infty} { R}(t)$.

To proceed further, one needs to know the explicit expression for the characteristic functional, $\Phi(t-t')$. 
For the random telegraph noise, $\Phi(t)$, obeys the following differential equation \cite{KV2}:
\begin{align}
\frac{d^2}{dt^2} \Phi(t) + 2\gamma \frac{d}{dt} \Phi(t) + D^2\sigma^2 \Phi(t) =0.
\end{align}
The solution is given by \cite{KV2,NB1},
\begin{align} \label{AP1a}
\Phi(t) = e^{-\gamma t}\Big (  \cosh\big(\sqrt{1- \mu^2}\,{\gamma t}\big)  + \frac{1}{\sqrt{1- \mu^2}}  \sinh\big(\sqrt{1- \mu^2}\,{\gamma t}\big) \Big ),
\end{align}
where $\mu = D\sigma/\gamma$.

\begin{figure}[tbh]
	\begin{center}
		\scalebox{0.325}{\includegraphics{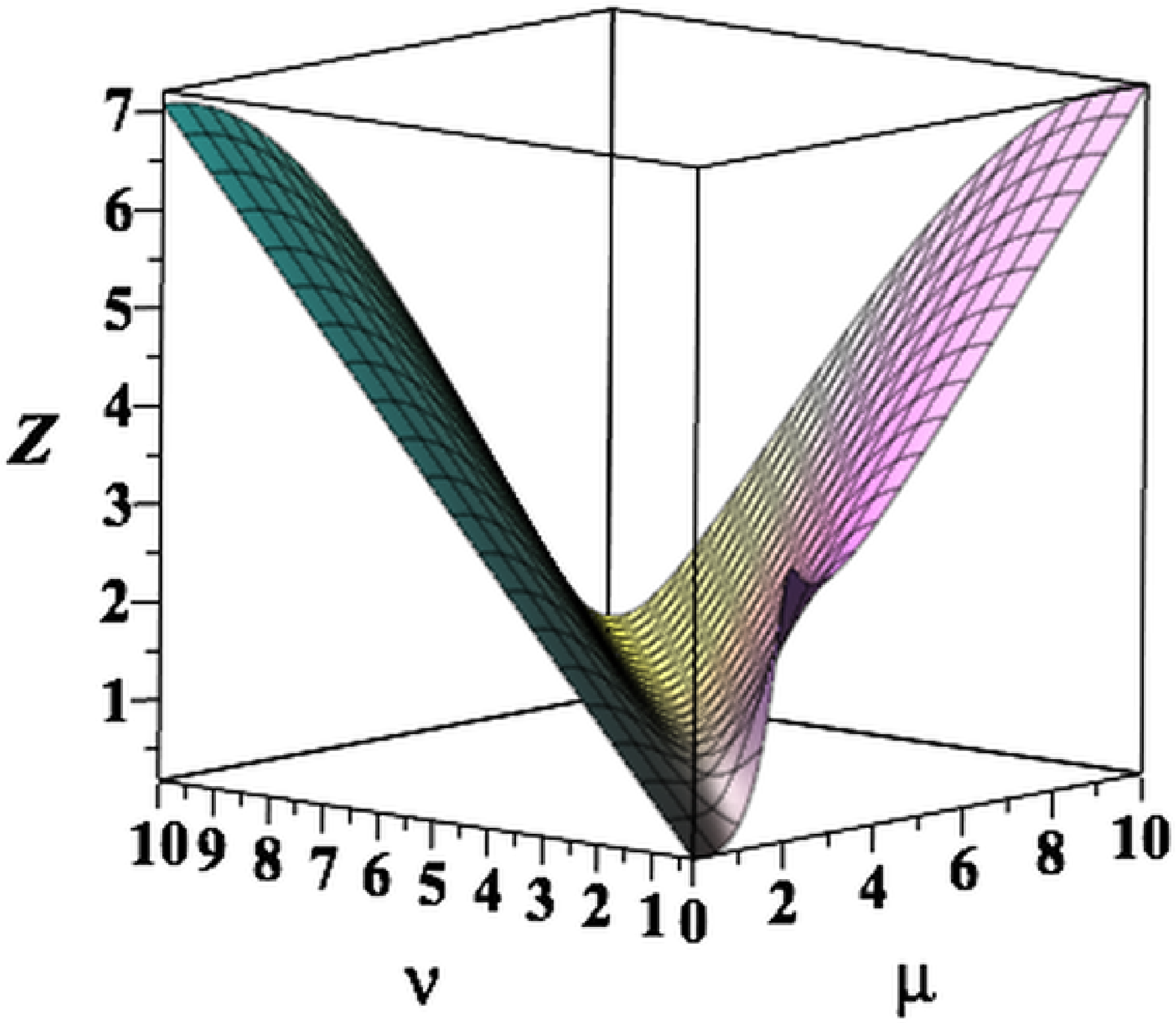}}
		(a)
		\scalebox{0.35}{\includegraphics{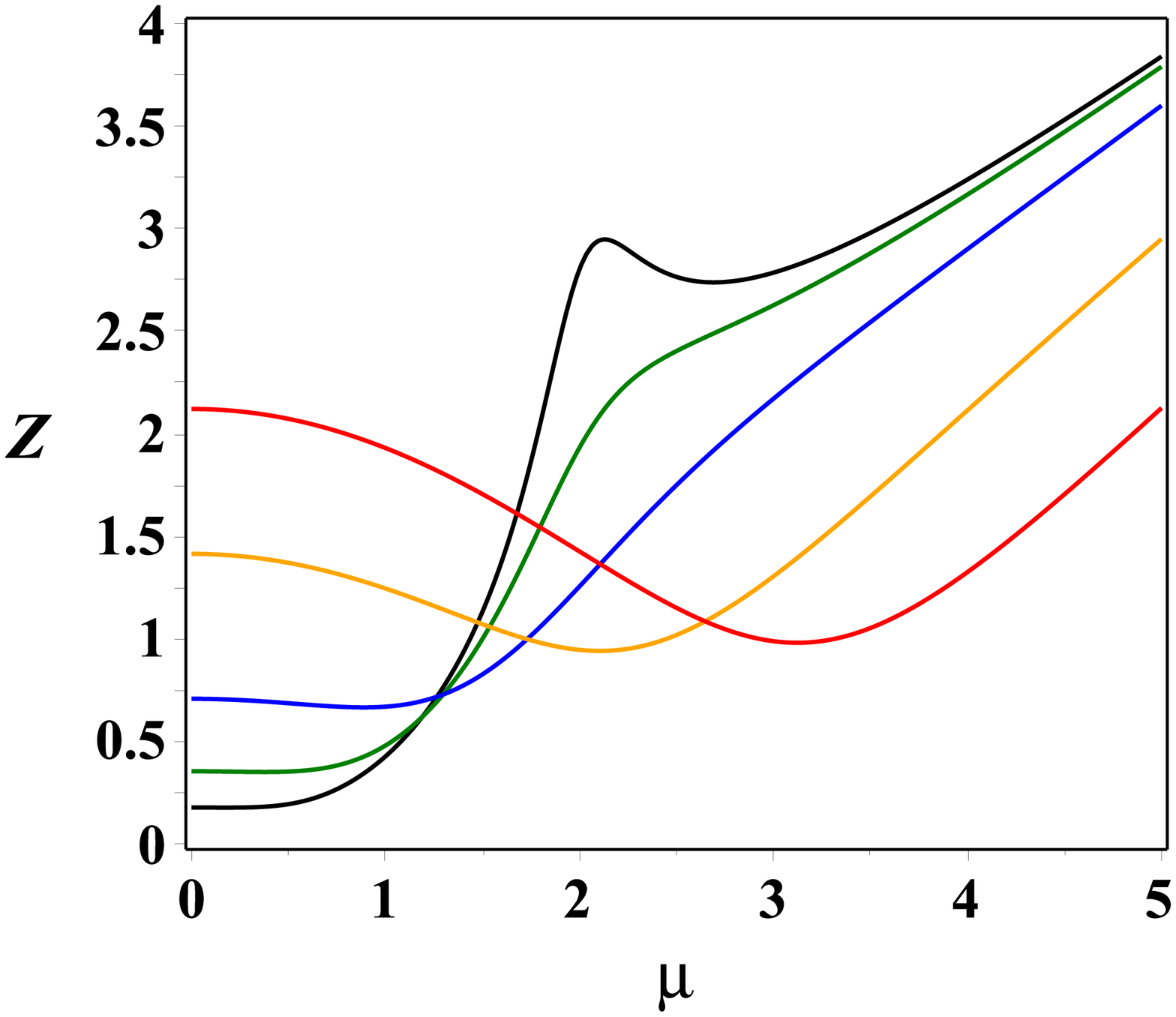}}
		(b)
	\end{center}
	\caption{(Color online)  Left panel. Dependence of $Z$ on $(\mu,\nu)$. Right panel. Dependence of $Z$ on $\mu$, for different values of $\nu$: $\nu=0.25$ (black curve), $\nu=0.5$ (green curve), $\nu=1$ (blue curve), $\nu=2$ (orange curve),  $\nu=3$ (red).
		\label{ZFig1}}
\end{figure}

\begin{figure}[tbh]
	{\scalebox{0.35}{\includegraphics{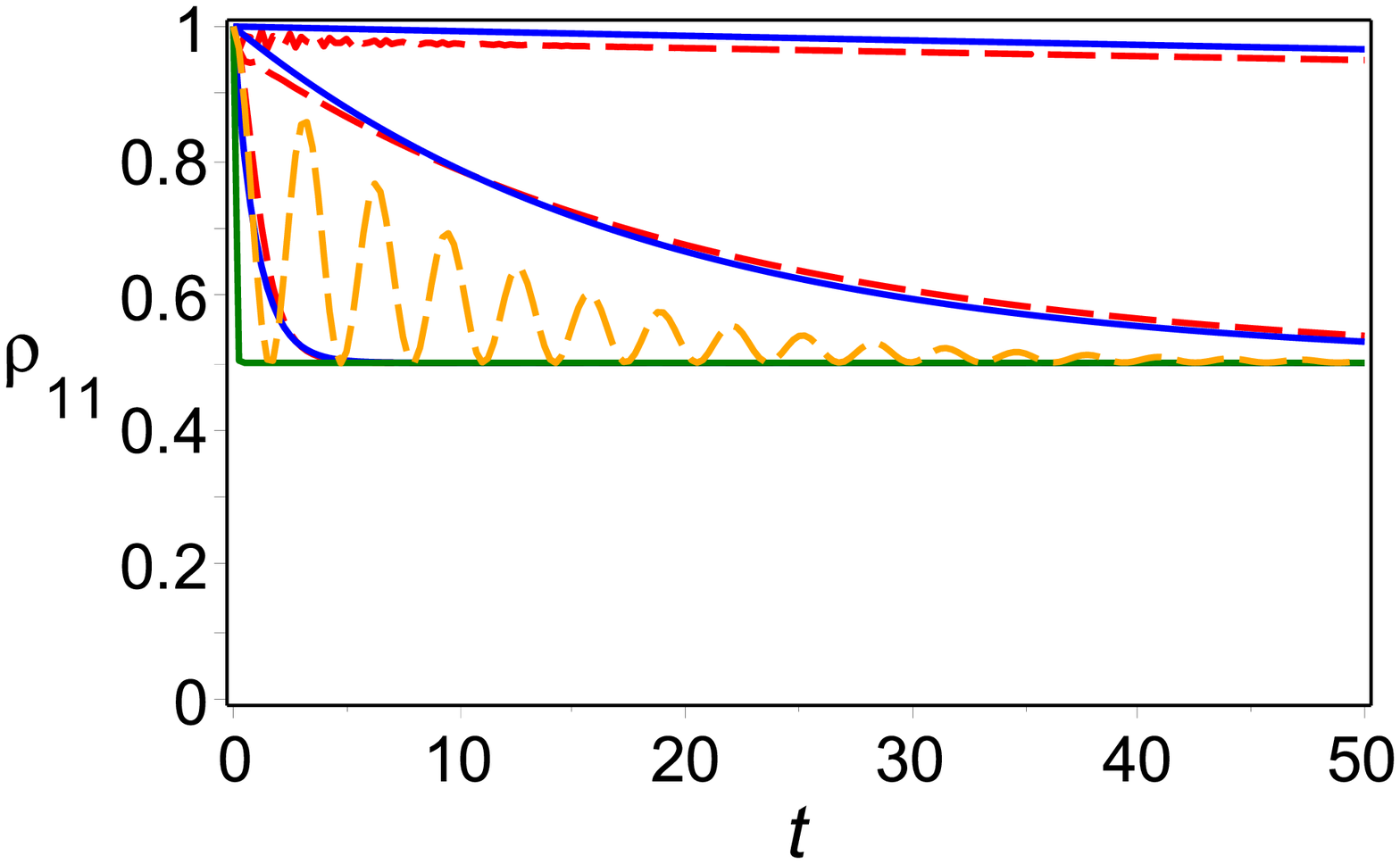}}
		(a)}
	{\scalebox{0.35}{\includegraphics{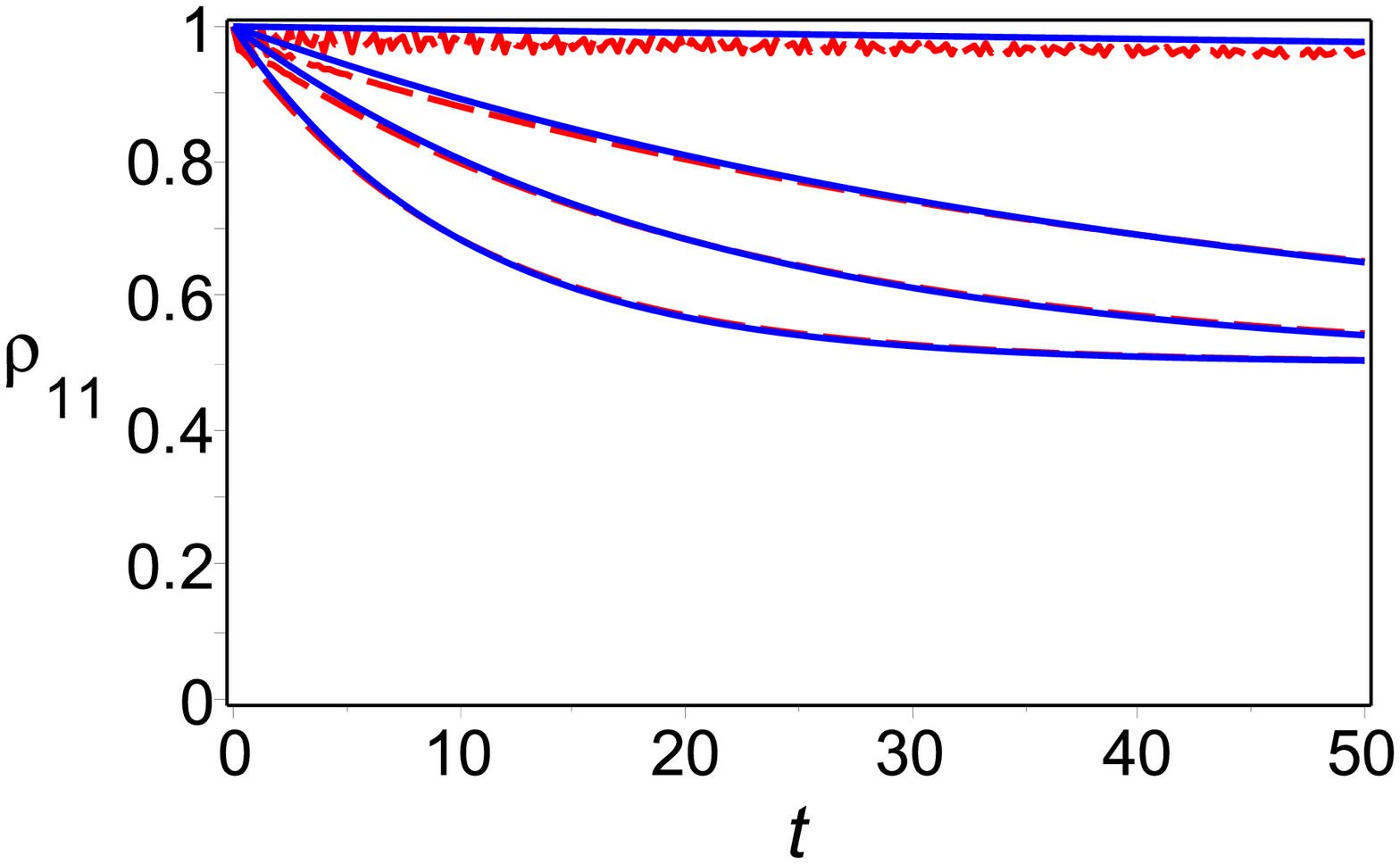}} 
		(b)}
	\scalebox{0.35}{\includegraphics{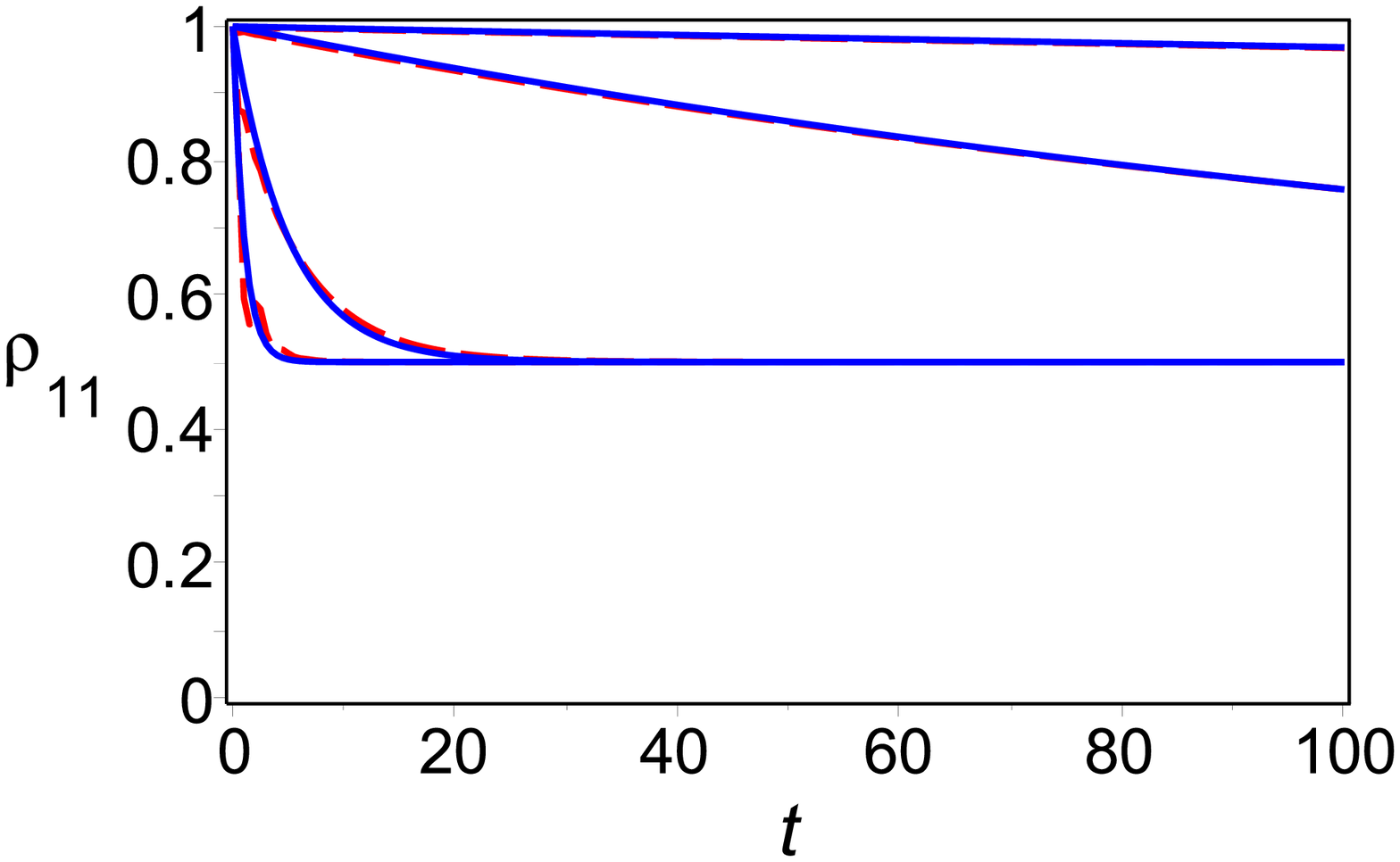}} 
	(c)
	\scalebox{0.35}{\includegraphics{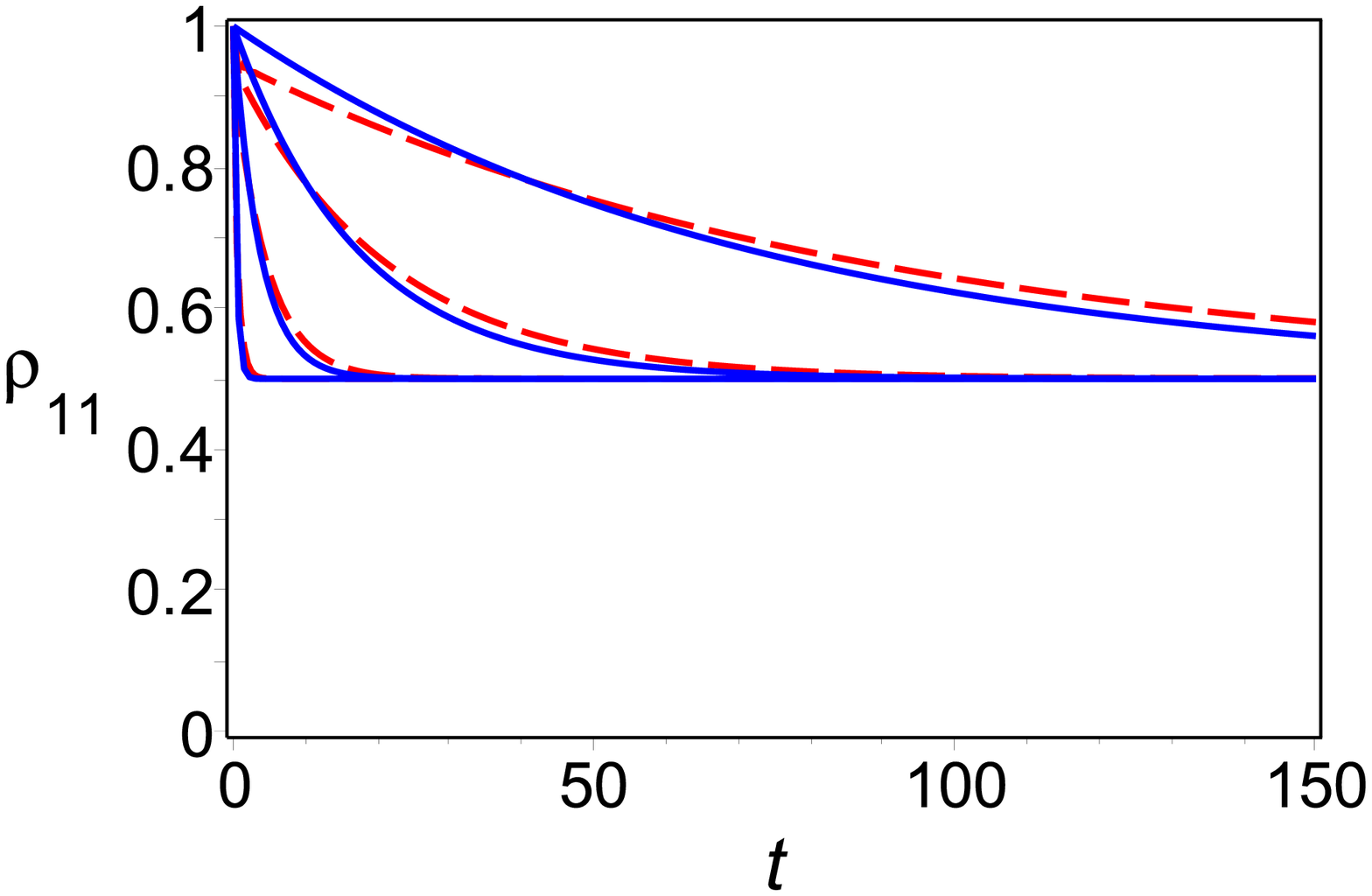}} 
	(d)
	\caption{(Color online)   Solution of Eq. (\ref{AS1}) (solid curves)  and Eqs. (\ref{AX5}) - (\ref{AX6c}) (dashed curves). From the top to the bottom: (a) $(\varepsilon,D,\gamma) =[(10,1,2),(10,5,2),(10,10,2)]$ (blue, red), $(\varepsilon,D,\gamma) = [(10,10,0.1)]$ (green, orange; here, the condition for applicability,  $V \ll\gamma$,  is violated); (b) $(\varepsilon,D,\gamma) =[(10,1,20),(10,5,20),(10,10,40), (10,10,20)]$; (c) $(\varepsilon,D,\gamma) =[(40,5,10),(20,5,10),(10,5,10),(5,5,10), (2,5,10)]$. In the cases (a) - (c), we chose, $V=\sigma =1$. (d) $(D,\gamma)=[(5,10),(10,10),(20,10), (40,10)]$, $V=10$,  $\varepsilon=60$, $\sigma = 1$.
		\label{FigG}}
\end{figure}

Further, it is convenient to define a complex kernel, $ {\tilde K }(t)=|V_{12}|^2\Phi(t) e^{-i\varepsilon t} $. From here it follows: ${ R}(t)  = 2\Re (\tilde{ R}(t) )$, where $\tilde{ R}(t) =\int_0^t{\tilde K }(s)ds$.  Employing (\ref{AP1}), we find,
\begin{align} \label{Ra1}
\tilde{ R}(t) =\frac{V^2}{\gamma}\int_{0}^{\gamma t}  e^{-(1+i\nu)\tau}\Big (  \cosh\big(\sqrt{1- \mu^2}\,{\tau}\big)  + \frac{1}{\sqrt{1- \mu^2}}  \sinh\big(\sqrt{1- \mu^2}\,{\tau}\big) \Big )d\tau
\end{align}
where $\nu =\varepsilon/\gamma$.  Performing the integration, we obtain,
\begin{align}\label{Ra2}
\tilde{R}(t) = &\frac{V^2}{\gamma(\mu^2-\nu^2 + 2i\nu)}\Big \{2+i\nu - e^{-(1 +i\nu)\gamma t}\Big( (2+i\nu) \cosh(\sqrt{1- \mu^2}\gamma t) \nonumber \\
&+ \big(\sqrt{1- \mu^2}+\frac{1+i\nu}{\sqrt{1- \mu^2}}\big) \sinh(\sqrt{1- \mu^2}\gamma t)\Big)\Big\},
\end{align}
where $\nu =\varepsilon/\gamma$. This yields,
\begin{align}\label{R2}
{  R}(t) =  &\Re\bigg \{\frac{2V^2}{\gamma(\mu^2-\nu^2 + 2i\nu)}\bigg (2+i\nu -  e^{-(1 +i\nu)\gamma t}\Big( (2+i\nu) \cosh(\sqrt{1- \mu^2}\gamma t) \nonumber \\
&+ \big(\sqrt{1- \mu^2}+\frac{1+i\nu}{\sqrt{1- \mu^2}}\big) \sinh(\sqrt{1- \mu^2}\gamma t)\Big)\bigg)\bigg\}.
\end{align}
Using Eq. (\ref{R2}), we obtain the asymptotic rate, $\Gamma$, 
\begin{align}\label{R2ab}
{\Gamma}=\stackrel{t\rightarrow \infty}{\longrightarrow}   \frac{8V^2\mu^2 }{\gamma((\mu^2-\nu^2 )^2+ 4\nu^2)}= \frac{8\gamma V^2 D^2\sigma^2 }{(D^2\sigma^2-\varepsilon^2 )^2+ 4\gamma^2\varepsilon^2}.
\end{align}
In two limiting cases, $D\sigma \ll \varepsilon$ (weak noise) and $D\sigma \gg \varepsilon$  (strong noise), we obtain,
\begin{align}\label{R2ac}
 {\Gamma} = \frac{8\gamma V^2  D^2\sigma^2 }{\varepsilon^2 (\varepsilon^2 + 4\gamma^2)}, \quad D\sigma \ll \varepsilon, \\
 {\Gamma} = \frac{8\gamma V^2 D^2\sigma^2 }{D^4\sigma^4+ 4\gamma^2\varepsilon^2}, \quad D\sigma \gg \varepsilon.
\end{align}

\subsection*{Conditions for validity of the approximation}

When the condition, $|\int_0^\infty \tau K(\tau)d\tau |\ll1$, is satisfied,  we can approximate Eqs. (\ref{DC4a}) and (\ref{DC4b}) by the following system of ordinary differential equations,
\begin{align} \label{AD7a}
\frac{d}{dt}{\langle{\rho}}_{11}(t) \rangle=&- { R}(t)  (\langle{\rho}_{11}(t)\rangle - \langle{\rho}_{22}(t) \rangle ) + iV_{21}\langle\rho_{12}(0)\rangle - iV_{12}\langle\rho_{21}(0)\rangle,\\
\frac{d}{dt}{\langle{\rho}}_{22}(t)\rangle  =& { R}(t) ( \langle{\rho}_{11}(t)\rangle - \langle{\rho}_{22}(t)\rangle  ) - iV_{21}\langle\rho_{12}(0)\rangle + iV_{12}\langle\rho_{21}(0)\rangle,
\label{AD7b}
\end{align}
One can show that the condition for validity of this approximation can be written as:
\begin{align}\label{Z1}
\left |\Im\left (\frac{\partial}{\partial \varepsilon}{ \tilde\Gamma}\right )\right |\leq \left |\frac{\partial}{\partial \varepsilon}{ \tilde\Gamma}\right | \ll 1,
\end{align}
where $\tilde \Gamma =2\int^{\infty}_0 t{\tilde K}(t) dt$. Introducing the dimensionless parameters,  $a=V/\gamma$ and $\nu =\varepsilon/\gamma$, we rewrite Eq. (\ref{Z1}) as, $a^2\ll \min Z $, where $Z=a^2/|\partial\tilde \Gamma/\partial\nu|$.

To find the function, $Z$, we use Eq. (\ref{R2}), which gives,
\begin{align}\label{G2}
\tilde{ \Gamma} = &\frac{2  |V_{12}|^2 }{\gamma}\cdot\frac{2\mu^2 +i\nu( (\mu^2-\nu^2) -4 )}{(\mu^2-\nu^2)^2 + 4\nu^2}.
\end{align}
After some algebra we obtain,
\begin{align}
Z=\frac{(\mu^2-\nu^2)^2 + 4\nu^2}{2\sqrt{16\nu^2 +(\mu^2 + \nu^2 -4)^2}}.
\end{align}

In Fig. \ref{ZFig1}, the function,  $Z(\mu,\nu)$, is shown. As one can see, the minimum of the function, $Z(\mu,\nu)$, is achieved when, $\mu\approx \nu$. Using this result, we obtain the following estimate: 

\begin{align}\label{est}
\frac{|V_{12}|}{\gamma} \ll \frac{\sqrt{2}\,\nu}{(\nu^4 + 4)^{1/4}}.
\end{align}

It follows from (\ref{est}) (see also Fig. \ref{ZFig1}b), that for $\nu \geq1$ the condition for validity of the approximation, leading to the differential equations (\ref{AD7a}) and (\ref{AD7b}), can be roughly estimated as: $V\ll \gamma$. 

In Fig. \ref{FigG}, we compare the numerical solutions of the approximate  Eq. (\ref{AS1}) (dashed curves)  with the corresponding solutions of the exact Eqs. (\ref{AX6}) - (\ref{AX6e}) (solid curves).  When $V\lesssim \gamma$, one can observe good agreement between both solutions.  However, when the condition of applicability, $V \ll\gamma$,  is violated,  one has disagreement between the approximate and the exact solutions. (See green and orange curves in Fig. \ref{FigG}a.)

\begin{figure}[tbh]
	\scalebox{0.325}{\includegraphics{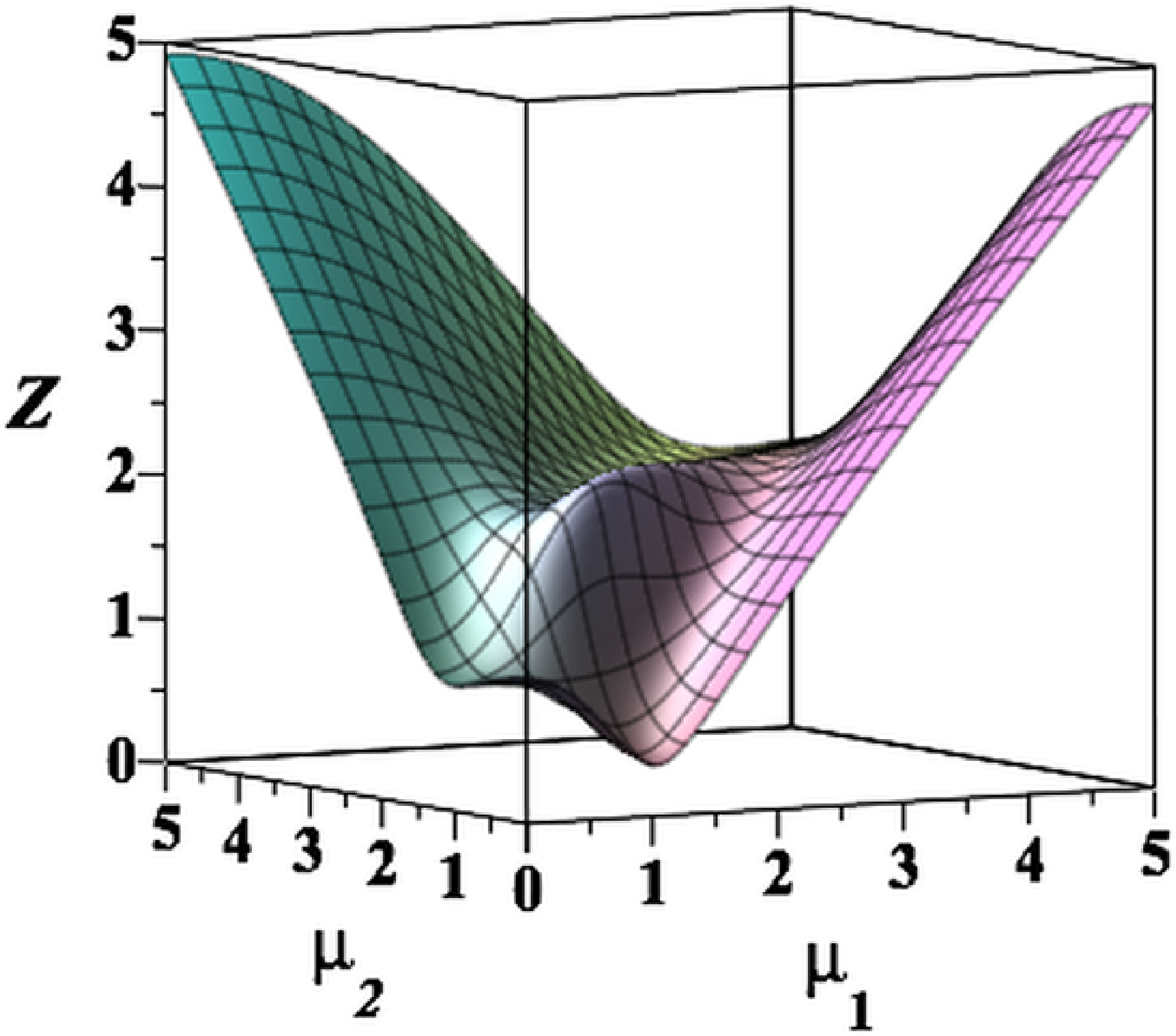}}
	(a)
	\scalebox{0.325}{\includegraphics{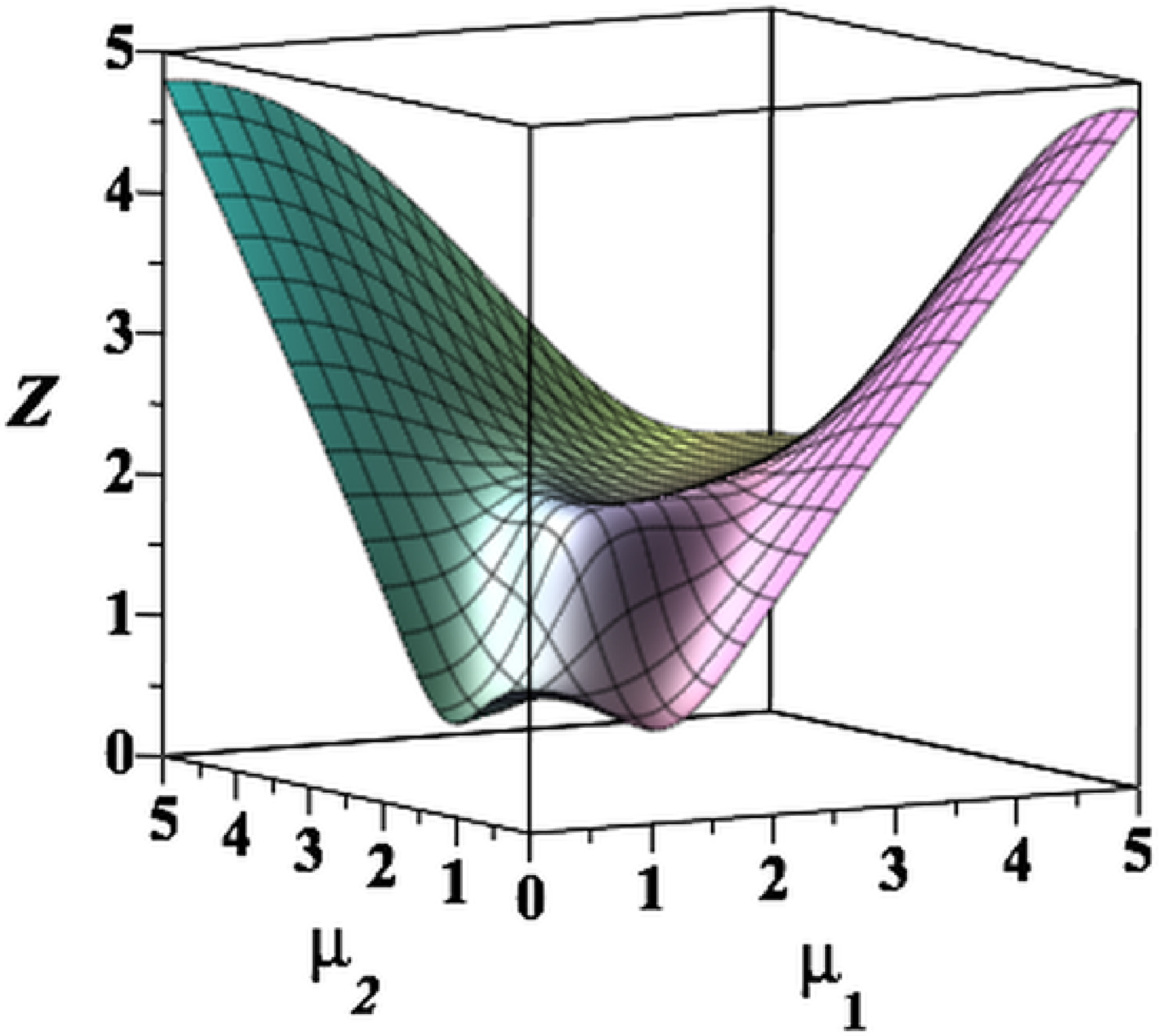}}
	(b)
	\caption{(Color online)  Dependence of $Z$ on $(\mu_1,\mu_2)$ ($\nu =1$).  Left panel: $g_1=0.25$, $g_2=0.75$. Right panel: $g_1=g_2=0.5$. 
		\label{ZFig2}}
\end{figure}

\section*{Two uncorrelated noises}

In the case of two uncorrelated RTPs, the behavior of the system can be described by the same approximate system of differential equation, as in the case of a single noise,
\begin{align} \label{BD7a}
\frac{d}{dt}{\langle{\rho}}_{11}(t) \rangle=&- { R}(t)  (\langle{\rho}_{11}(t)\rangle - \langle{\rho}_{22}(t) \rangle ) + iV_{21}\langle\rho_{12}(0)\rangle - iV_{12}\langle\rho_{21}(0)\rangle,\\
\frac{d}{dt}{\langle{\rho}}_{22}(t)\rangle  =& { R}(t) ( \langle{\rho}_{11}(t)\rangle - \langle{\rho}_{22}(t)\rangle  ) - iV_{21}\langle\rho_{12}(0)\rangle + iV_{12}\langle\rho_{21}(0)\rangle,
\label{BD7b}
\end{align}
where, ${ R}(t) =\int_0^t  K(\tau)d\tau$, and, 
\begin{align}\label{K1}
  K(t-t') =  2|V_{12}|^2\Phi(t-t') \cos (\varepsilon(t-t')).
\end{align}
The only difference is that now the characteristic functional, $\Phi(t)$, is presented as a product: $\Phi(t) =\Phi_1(t)\Phi_2(t)$.  The characteristic  functional, $\Phi_a(t)$, of each independent RTP, is given by \cite{BGA,NB1},
\begin{align} \label{APa1}
\Phi_a(t) = e^{-\gamma_a t}\Big (  \cosh\big(\sqrt{\gamma^2_a- d_a^2}\,{ t}\big)  + \frac{1}{\sqrt{\gamma^2_a- d_a^2}}  \sinh\big(\sqrt{\gamma^2_a- d_a^2}\,{ t}\big) \Big ), \quad a =1,2,
\end{align}
where, $d_a = (\lambda^{(a)}_1- \lambda^{(a)}_2 )\sigma_a$, denotes the amplitude of the $a$-th noise.

Computation of the asymptotic rate, $\Gamma=2\lim_{t\rightarrow \infty} R(t)$, yields,
\begin{align}
\Gamma=&\displaystyle\frac{2|V_{12}|^2}{\alpha _{1}\,\alpha _{2}\, (\gamma_1 + \gamma_2)}\Re\Bigg(\displaystyle \,{\frac { \left( g_{1}\,g_{2}-\alpha _{1}\,\alpha _{2} \right)  \left( 1+i\nu \right) + \left( \alpha _{1}\,g_{2} -g_{1}\,\alpha _{2}\right)  \left( \alpha _{1}-\alpha _{2} \right) }{ \left( \alpha _{1}-\alpha _{2} \right) ^{2}- \left( 1+i\nu \right) ^{2}  }} \nonumber \\
&- \displaystyle \,{\frac { \left( g_{1}\,g_{2}+\alpha _{1}\,\alpha _{2} \right)  \left( 1+i\nu \right) + \left( g_{1}\,\alpha _{2}+\alpha _{1}\,g_{2} \right)  \left( \alpha _{1}+\alpha _{2} \right) }{\left( \alpha _{1}+\alpha _{2} \right) ^{2}- \left( 1+i\nu \right) ^{2}  }}\Bigg),
\label{Ga1}
\end{align}
where
\begin{align}
&\alpha_1= \sqrt{g_1^2 -\mu_1^2}, \quad \alpha_2= \sqrt{g_2^2 -\mu_2^2}, \quad g_1= \frac{\gamma_1 }{\gamma_1 + \gamma_2}, \quad g_2= \frac{\gamma_2 }{\gamma_1 + \gamma_2}, \nonumber \\
& \displaystyle \mu_1=\frac{d^1_1 -d^1_2}{\gamma_1 + \gamma_2}, \quad \mu_2=\frac{d^2_1 -d^2_2}{\gamma_1 + \gamma_2} .
\end{align}

The condition for validity for approximating the integro-differential equations by the system of differential equations is the same as in the case of a single noise: $|\int_0^\infty \tau K(\tau)d\tau |\ll1$. Performing the same procedures, as in the case of a single noise, we obtain,
\begin{align}
\frac{|V_{12}|}{\gamma_1 + \gamma_2} \ll\min Z,
\end{align}
 where $Z=a^2/|\partial\tilde \Gamma/\partial\nu|$, and
 \begin{align}
\tilde\Gamma=&\displaystyle\frac{1}{\alpha _{1}\,\alpha _{2}}\Bigg(\displaystyle \,{\frac { \left( g_{1}\,g_{2}-\alpha _{1}\,\alpha _{2} \right)  \left( 1+i\nu \right) + \left( \alpha _{1}\,g_{2} -g_{1}\,\alpha _{2}\right)  \left( \alpha _{1}-\alpha _{2} \right) }{ \left( \alpha _{1}-\alpha _{2} \right) ^{2}- \left( 1+i\nu \right) ^{2}  }} \nonumber \\
&- \displaystyle \,{\frac { \left( g_{1}\,g_{2}+\alpha _{1}\,\alpha _{2} \right)  \left( 1+i\nu \right) + \left( g_{1}\,\alpha _{2}+\alpha _{1}\,g_{2} \right)  \left( \alpha _{1}+\alpha _{2} \right) }{\left( \alpha _{1}+\alpha _{2} \right) ^{2}- \left( 1+i\nu \right) ^{2}  }}\Bigg),
\label{Gb1}
\end{align}

\begin{figure}[tbh]
{\scalebox{0.325}{\includegraphics{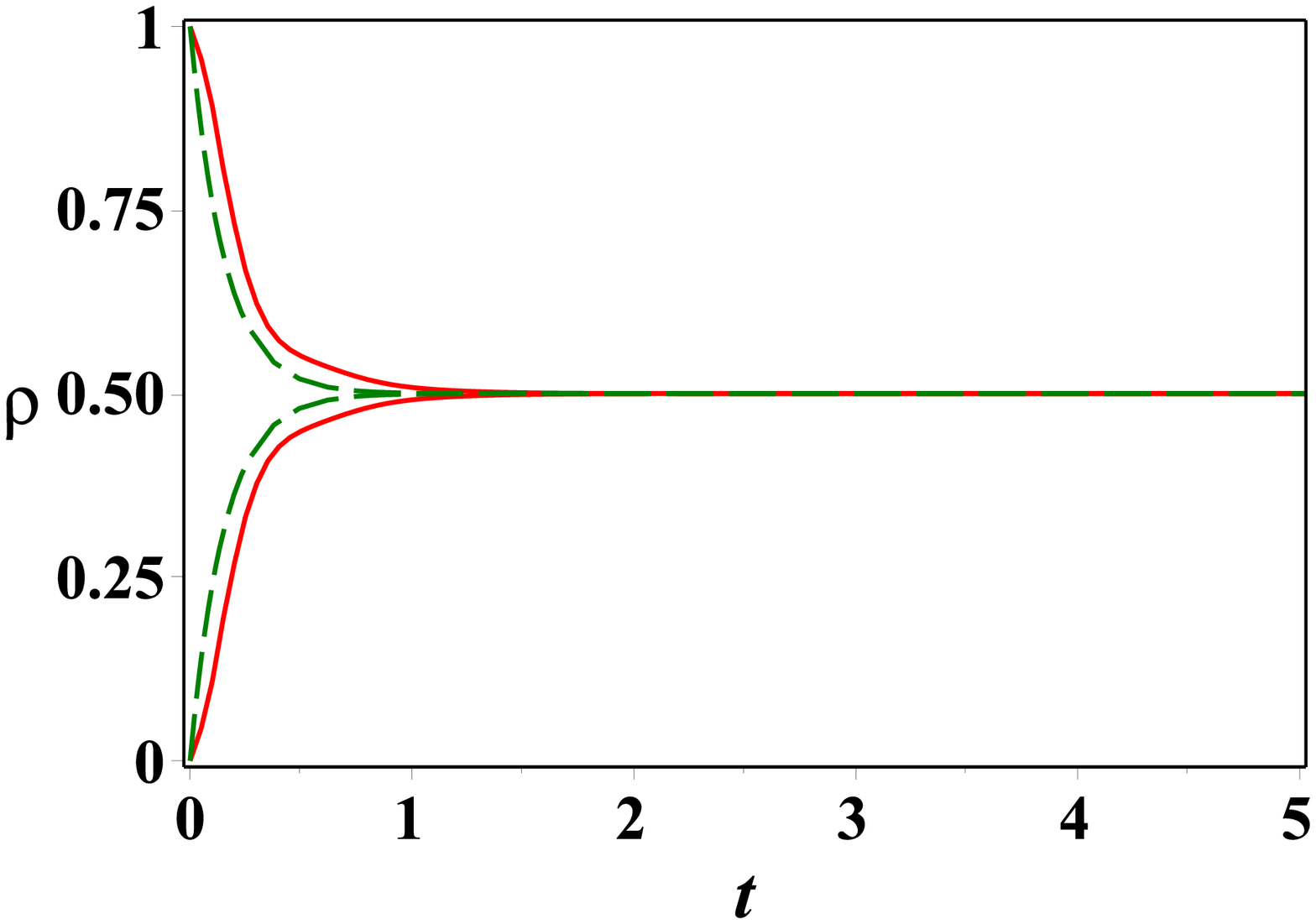}}
(a)}
{\scalebox{0.35}{\includegraphics{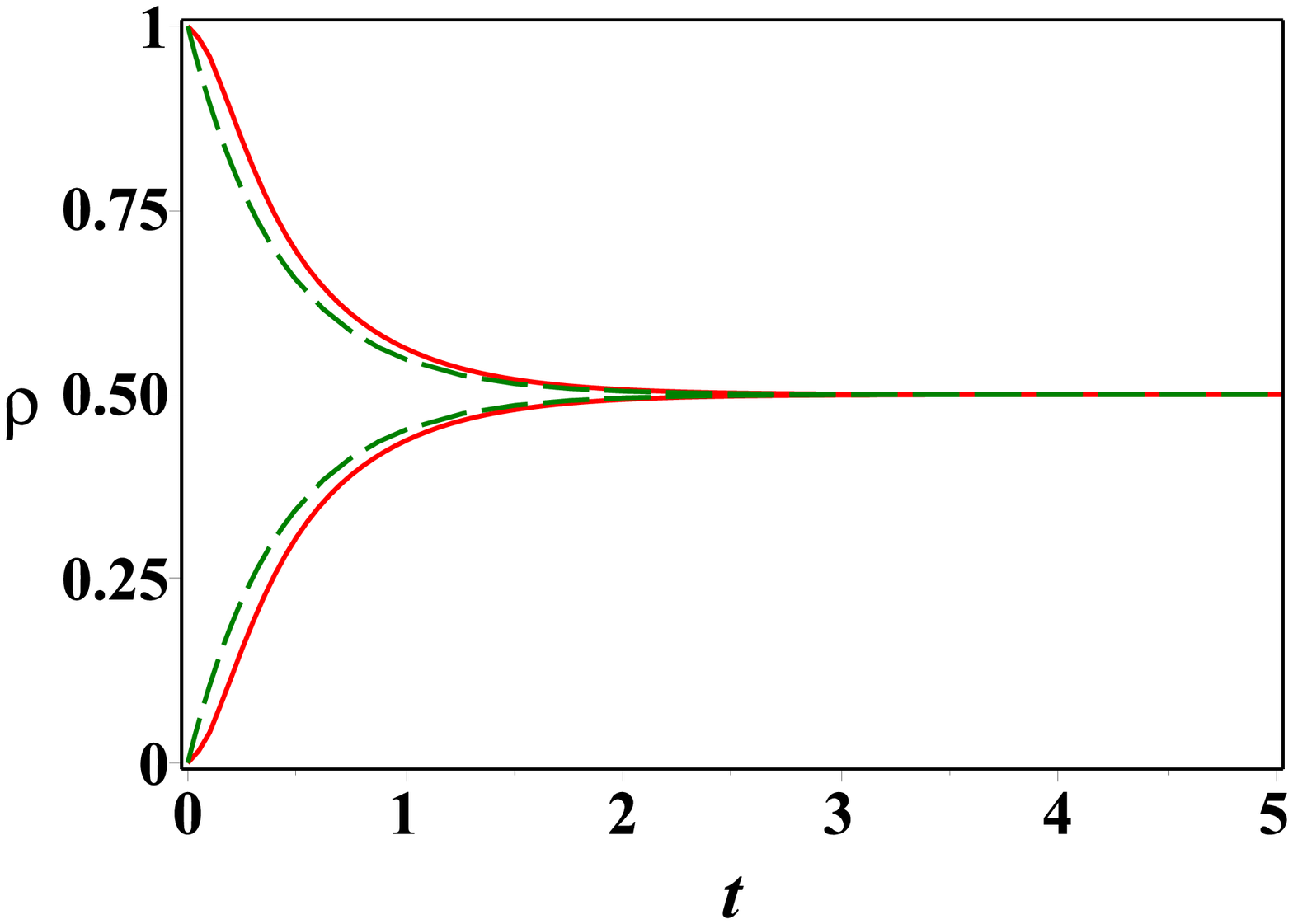}} 
(b)}
\scalebox{0.35}{\includegraphics{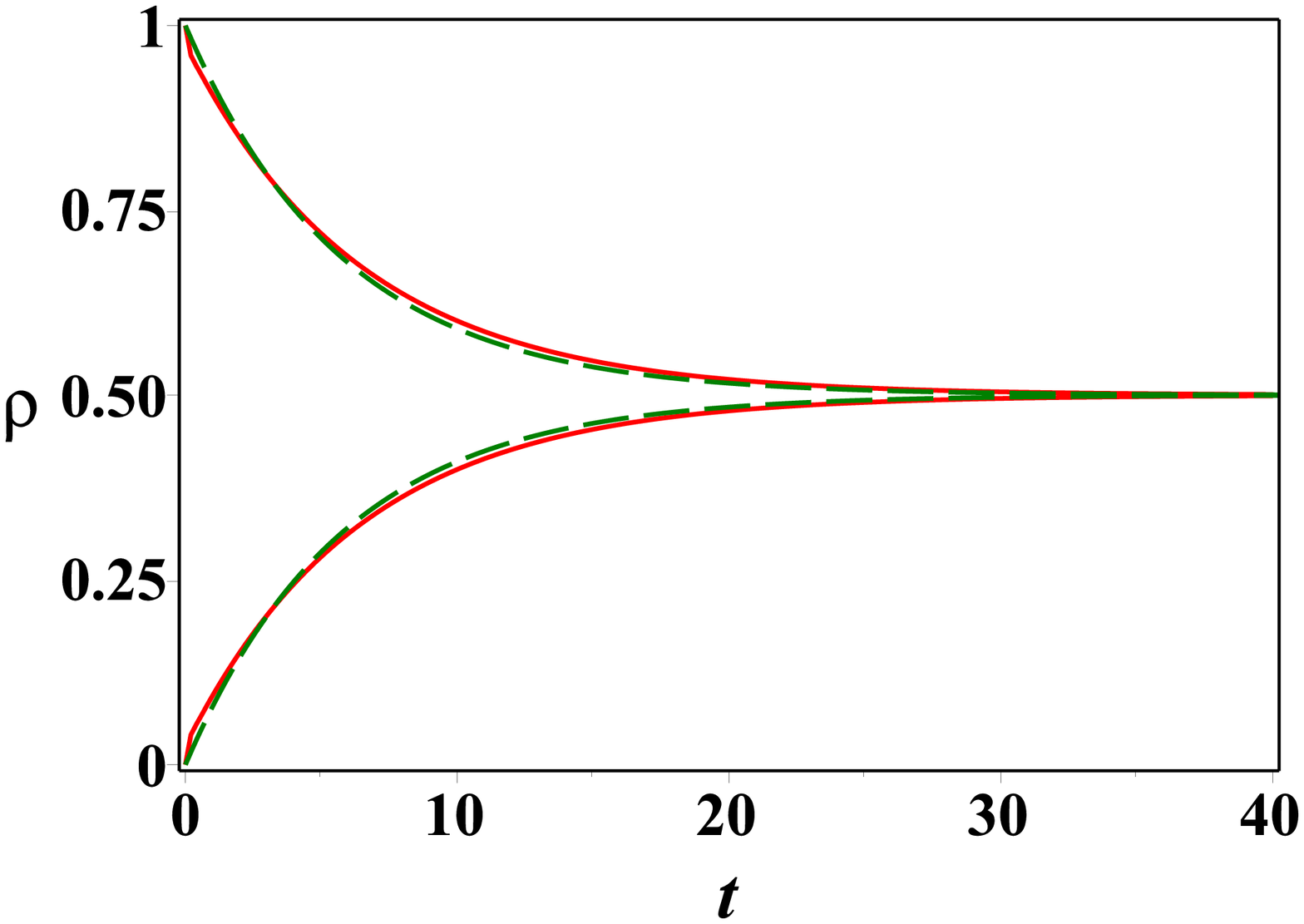}} 
(c)
\scalebox{0.35}{\includegraphics{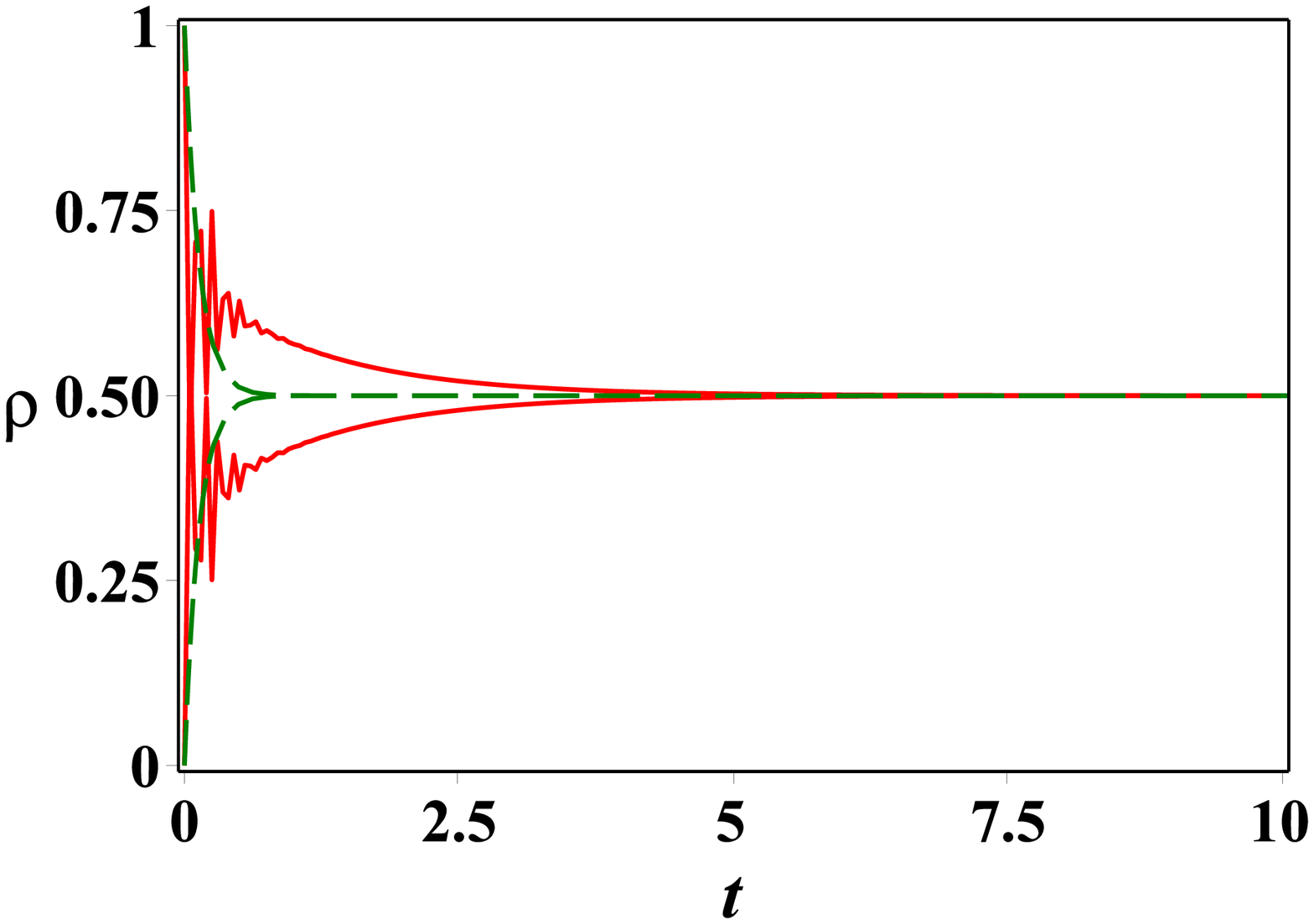}} 
(d)
\caption{(Color online)  Dashed curves: solution of the approximate Eq. (\ref{C6b}). Solid curves: solutions of the exact system of differential equations, for two uncorrelated RTPs. Parameters: $\varepsilon= 30,$,  $\gamma_1=10$, $\gamma_2=15$. (a)  $V_{12}=5$, $d^{(1)}_1= 30$, $d^{(1)}_2= 0$,  $d^{(2)}_1= 0$, $d^{(2)}_2= 10$, (b) $V_{12}=3$, $d^{(1)}_1= 30$, $d^{(1)}_2= 0$,  $d^{(2)}_1= 0$, $d^{(2)}_2= 10$, (c)  $V_{12}=3$, $d^{(1)}_1= 10$, $d^{(1)}_2= 0$,  $d^{(2)}_1= 0$, $d^{(2)}_2= 10$, (d) $V_{12}=20$, $d^{(1)}_1= 30$, $d^{(1)}_2= 0$,  $d^{(2)}_1= 0$, $d^{(2)}_2= 10$.
Initial conditions:  $\rho_{11}(0) =1$,  $\rho_{22}(0) =0$.
\label{FigG2}}
\end{figure}

In Fig. \ref{ZFig2}, the function,  $Z(\mu_1,\mu_2)$, is shown. As one can see, the minimum of the function, $Z(\mu,\nu)$, occurs when $\mu_1\approx \mu_2$. It follows from this (see also Fig. \ref{ZFig1}b) that for, $\nu \geq1$, the condition for validity of the approximation, leading to the differential equations (\ref{AD7a}) and (\ref{AD7b}), can be roughly estimated as: $V\ll \gamma_1+\gamma_2$. 

In Fig. \ref{FigG2}, we compare the numerical solutions (dashed curves)  of the approximate  equation,
\begin{align}\label{C6b}
{\langle{\rho}}_{11}(t)\rangle = \frac{1}{2} +   \frac{1}{2} e^{-\Gamma t},
\end{align}
 with the corresponding solutions (solid curves) of the exact Eqs. (8) - (10) from the main text of our paper.  When $V\lesssim \gamma_1 + \gamma_2$, one can observe a good agreement between the exact and the approximate solutions. However, when the condition of applicability, $V \ll\gamma_1 +\gamma_2$,  is violated,  one has disagreement between the two solutions. (See green and red curves in Fig. \ref{FigG2}d.)
\end{widetext}

\end{document}